\documentclass[a4paper,11pt]{article}
\pdfoutput=1 

\usepackage{jheppub} 

\usepackage[T1]{fontenc} 

\usepackage{amsmath}
\usepackage{graphicx}
\usepackage{dcolumn}
\usepackage{bm}
\usepackage{wasysym}
\usepackage{verbatim}
\usepackage{color}
\usepackage{mathtools}

\newcommand{\Beq}{\begin{equation}\begin{aligned}}
\newcommand{\Eeq}{\end{aligned}\end{equation}}
\newcommand{\be}{\begin{eqnarray}}
\newcommand{\ee}{\end{eqnarray}}
\def\p{\partial}

\newcommand{\mH}{\mathcal{H}}

\newcommand{\xp}{\xi_{A}}

\newcommand{\an}{\quad \textmd{and} \quad}
\newcommand{\where}{\quad \textmd{where} \quad}
\newcommand{\x}{\tilde{\tau}}

\newcommand{\ga}{g_{\!_{A}}}
\newcommand{\hgz}{\varphi}
\newcommand{\Hgz}{\boldsymbol{\varphi}}

\newcommand{\vx}{\vec{x}}

\newcommand{\mJ}{\mathcal{J}}
\newcommand{\kl}{\kappa_{\lambda}}
\newcommand{\bk}{\boldsymbol{k}}
\newcommand{\bx}{\boldsymbol{x}}

\title{\boldmath Schwinger Effect by an $SU(2)$ Gauge Field during Inflation}

\author[a]{K. D. Lozanov,}
\author[a]{A. Maleknejad,}
\author[a,b]{and E. Komatsu}

\affiliation[a]{Max-Planck-Institute for Astrophysics, Karl-Schwarzschild-Str. 1, 85741 Garching, Germany}
\affiliation[b]{Kavli Institute for the Physics and Mathematics of the Universe (Kavli IPMU,
WPI), UTIAS, The University of Tokyo, Chiba, 277-8583, Japan}

\emailAdd{klozanov@MPA-Garching.MPG.DE}
\emailAdd{amalek@MPA-Garching.MPG.DE}
\emailAdd{komatsu@MPA-Garching.MPG.DE}

\abstract{Non-Abelian gauge fields may exist during inflation. We study the Schwinger effect by an $SU(2)$ gauge field coupled to a charged scalar doublet in a (quasi) de Sitter background and the possible backreaction of the generated charged particles on the homogeneous dynamics. Contrary to the Abelian $U(1)$ case, we find that both the Schwinger pair production and the induced current decrease as the interaction strength increases. The reason for this suppression is the isotropic vacuum expectation value of the $SU(2)$ field which generates a (three times) greater effective mass for the scalar field than the $U(1)$. In the weak interaction limit, the above effect is negligible and both the $SU(2)$ and $U(1)$ cases exhibit a linear increase of the current and a constant conductivity with the interaction strength. We conclude that the Schwinger effect does not pose a threat to the dynamics of inflationary models involving an $SU(2)$ gauge field.}

\begin{document}
\maketitle
\flushbottom


\section{Introduction}
\label{sec:Intro}

A strong classical field can provide enough energy for the pair production of particles out of the vacuum. The Schwinger effect, the pair production by a static electric field, is one such example \cite{PhysRev.82.664}. In a curved spacetime, the effect receives contributions from the gravitational field as well. Recently, several aspects of the Schwinger effect in a (quasi) de Sitter background have been studied for scalar and fermionic matter \cite{Kobayashi:2014zza,Frob:2014zka,Hayashinaka:2016qqn,Sharma:2017ivh}. In these studies, the background field is a static and homogeneous $U(1)$ gauge field with a preferred direction which slightly breaks the spatial isotropy. It has been shown that in this setup, the Schwinger effect leads to a sizable particle production and backreaction in the strong field regime. 

In this work, we study the Schwinger effect by an $SU(2)$ gauge field coupled to a massive scalar doublet, $\boldsymbol{\varphi}$, in a (quasi) de Sitter background. It couples to the gauge fields via the covariant derivative term, $\left({\bf{D}}_{\mu}\boldsymbol{\varphi}\right)^{\dagger}{\bf{D}}^{\mu}\boldsymbol{\varphi}\,$. Such an interaction is well-motivated since it appears in the Standard Model of elementary particles and fields, although our $SU(2)$ and scalar fields are different from those in the Standard Model. We shall study the Schwinger production of scalar particles in an inflating universe and check if their energy is comparable to the background energy of the gauge fields. If it is, one expects the homogeneous gauge field modes to disappear quickly due to backreaction (the excited charged scalar field effectively screening the $SU(2)$ background). 

The Schwinger particle production was originally studied in the context of charged fermions \cite{PhysRev.82.664} in a non-zero gauge field background. We consider a bosonic field instead, since we expect more particle production (due to the absence of Pauli suppression of occupation numbers) and therefore a greater possibility to have backreaction. In other words, if the Schwinger excitation of a bosonic field does not lead to sizable backreaction effects, we do not anticipate to see them in a fermionic model either. 

Why study $SU(2)$ fields in a de Sitter background? As was first discovered by one of the authors (A.M.) \cite{Maleknejad:2011jw,Maleknejad:2011sq} and the subsequent works \cite{Adshead:2012kp,Adshead:2013nka,Adshead:2016omu,Adshead:2017hnc,Dimastrogiovanni:2012st,Dimastrogiovanni:2012ew, Namba:2013kia}, inflation with $SU(2)$ gauge fields shows rich phenomenology that is absent in the usual single-scalar-field inflation models. An important example is the generation of stochastic backgrounds of {\it chiral} gravitational waves (GWs) \cite{Adshead:2013nka,Dimastrogiovanni:2012ew,Maleknejad:2016qjz}. In particular, some gauge field models can generate observable primordial GWs either as a large field model or by sourced perturbations through their interaction with the gauge fields. Moreover, the mixing between the $SU(2)$ gauge field and perturbations in the scalar and tensor sectors are at the linear order and coming from different fluctuations. Hence, unlike the $U(1)$ gauge field, the enhancement of GWs and the modification in the scalar perturbations are uncorrelated at tree level. These setups are therefore able to generate detectable (and chiral) GWs \cite{Dimastrogiovanni:2016fuu} and tensor non-Gaussianity \cite{Agrawal:2017awz, Agrawal:2018mrg} while having little sourced scalar power spectrum and non-Gaussinities. Finally, having a non-zero source of parity violation during inflation, these models provide a natural setting for inflationary leptogenesis and explain the observed matter asymmetry in the Universe \cite{Maleknejad:2016dci, Caldwell:2017chz, Adshead:2017znw}. \footnote{This effect is not efficient enough for inflationary models involving $U(1)$ gauge fields \cite{Papageorgiou:2017yup}. However, a recent study \cite{Adshead:2018doq} of the subsequent post-inflationary dynamics in such models shows that successful baryogenesis can occur during preheating.} While the details of the models can be different, there are several things in common between these models. Among them is an (almost constant) isotropic and homogeneous field configuration for the $SU(2)$ gauge field. That gauge field vacuum expectation value (VEV) then generates almost constant electric and magnetic fields. We emphasise that the $su(2)$ algebraic isomorphy with $so(3)$ makes it possible to have a gauge field background which respects the spatial isotropy.

If backreaction on the $SU(2)$-background by the Schwinger effect is important, the copious generation of large-scale GWs can be suppressed. This can provide additional constraints in parameter space since regions with backreaction will not be relevant if GWs are observed. The dynamics of an isolated axion-$SU(2)$-spectator sector (only gravitationally coupled to other fields) already constrains the parameters of the system. The combination of the $SU(2)$ coupling constant, $g_{\!A}$, its background value, $Q$, and the Hubble parameter during inflation, $H$, $\xp \equiv g_{\!A}Q/H$, is bound to be $\sqrt{2}<\xp\lesssim5$, whereas $\epsilon_B\equiv \xp^4H^2/(g_{\!A}^2m_{\rm{pl}}^2)>10^{-10}$ for models with $r_{\rm{vac}}>10^{-4}$, see Ref. \cite{Agrawal:2018mrg}. In this work, we study in detail the Schwinger effect in this class of models and address the issue of backreaction.

This paper is organized as follows. In Section \ref{sec:setup} we present our theory in which we study the Schwinger effect. In Section \ref{sec:SchwingerEffect} we study the evolution of a charged scalar field in the external $SU(2)$ background and the resulting Schwinger pair production. The generated scalar current is presented in Section \ref{sec:Indcurr}, and the details of the computations are given in Appendices \ref{Current-appendix} and \ref{Adiabatic Subtraction}. In Section \ref{sec:Constr} we compute the backreaction on the gauge field background. We conclude in Section \ref{sec:concl}.


\section{The Setup}
\label{sec:setup}

We consider a charged complex scalar doublet
\Beq
\label{eq:Higgs}
\boldsymbol{\varphi}=\frac{1}{\sqrt{2}}\begin{pmatrix}\varphi^{1}\\ \varphi^{2}\end{pmatrix}\,,
\Eeq
which is coupled to a non-Abelian gauge field with the action
\Beq
\label{eq:FullAction}
S_{\rm{matter}}=\int\text{d}^4x\sqrt{-g}\left[\left({\bf{D}}_{\mu}\boldsymbol{\varphi}\right)^{\dagger}{\bf{D}}^{\mu}\boldsymbol{\varphi}-m^2\boldsymbol{\varphi}^{\dagger}\boldsymbol{\varphi}\right]\,.
\Eeq
The covariant derivative is given as
\Beq
{{\bf{D}}}_{\mu}\boldsymbol{\varphi}=\left({\bf{I}}_{2\times2}\nabla_{\mu}+ig_{\!A}\bf{A}_{\mu}\right)\boldsymbol{\varphi}\,,
\Eeq
where ${\bf{I}}_{2\times2}$ is a $2\times2$ identity matrix. We write the $SU(2)$ gauge field as
\Beq
\label{eq:SU2}
{\bf{A}}_{\mu}=A^{a}_{\mu}\frac{\boldsymbol{\sigma}^a}{2},
\Eeq 
in which $\boldsymbol{\sigma}^a$ are the Pauli matrices.
The field equation of $\boldsymbol{\varphi}$ is
\Beq
\label{eq:EoMHiggsGeneral}
\Box\boldsymbol{\varphi}+2ig_{\!A}{\bf{A}}_{\mu}\partial^{\mu}\boldsymbol{\varphi}+ig_{\!A}(\nabla_{\mu}{\bf{A}^{\mu}})\boldsymbol{\varphi}-g_{\!A}^2{\bf{A}_{\mu}}{\bf{A}^{\mu}}\boldsymbol{\varphi}+m^2\boldsymbol{\varphi}=\boldsymbol{0}\,.
\Eeq
Here, we are interested in the Schwinger process for $\boldsymbol{\varphi}$ during inflation in the presence of a (almost) constant, isotropic and homogeneous $SU(2)$ gauge field. There are several inflationary models involving  $SU(2)$ gauge fields in the literature that can generate the desired gauge field configuration \cite{Maleknejad:2011jw,Maleknejad:2011sq,Adshead:2012kp,Adshead:2013nka,Maleknejad:2016qjz,Caldwell:2017chz,Adshead:2016omu,Dimastrogiovanni:2016fuu,Adshead:2017hnc}. In the temporal gauge with $A^a_0=0$, we have \cite{Maleknejad:2011jw}
\Beq
\label{eq:SU2Ansatz}
A^{a}_j=a(\tau)Q(\tau)\delta^a_j\,,
\Eeq
where $a(\tau)$ is the Friedmann-Robertson-Walker scale factor, $Q(\tau)$ behaves as a scalar under spatial rotations and $\delta^a_j$ is the Kronecker delta. In this paper, we choose to work with one particular example of inflationary models involving $SU(2)$ gauge fields to constrain the model by the Schwinger effect. However, we emphasise that our conclusions are generic for models with the isotropic and homogeneous background gauge field configuration. 

The inflationary model that we choose is specified by the action of the spectator model studied in \cite{Dimastrogiovanni:2016fuu}
\Beq
\label{eq:Action}
S_{\rm{Inf}}=S_{\rm{EH}}+S_{\phi}+S_{\rm{spec}}\,,
\Eeq
where $S_{\rm{EH}}$ and $S_{\phi}$ are the Einstein-Hilbert and the inflaton actions, respectively, which are responsible for the inflation of the universe \cite{Guth:1980zm,Sato:1980yn,Linde:1981mu,Albrecht:1982wi}. The axion, $\chi$, and the gauge field act as spectator fields given by the action 
\Beq
\label{the-spec}
S_{\rm{spec}}=\int\text{d}^4x\sqrt{-g}\left[\frac{1}{2}\partial_{\mu}\chi\partial^{\mu}\chi-V(\chi)-\frac{1}{2}{\bf{Tr}(\boldsymbol{F}_{\mu\nu}\boldsymbol{F}^{\mu\nu})}-\frac{\lambda_{\!A}\chi}{2f}{\bf{Tr}(\boldsymbol{F}_{\mu\nu}\tilde{\boldsymbol{F}}^{\mu\nu})}\right]\,,
\Eeq
where $V(\chi)$ is the axion potential, $f$ is the decay constant and $\bf{F_{\mu\nu}}$ is the field strength tensor of the $SU(2)$ gauge field
\Beq
{\bf{F_{\mu\nu}}}=\p_{\mu}{\bf{A}_{\nu}}-\p_{\nu}{\bf{A}_{\mu}}+ig_{\!A}\left({\bf{A}_{\mu}}{\bf{A}_{\nu}}-{\bf{A}_{\nu}}{\bf{A}_{\mu}}\right)\,.
\Eeq
The last term in Eq. \eqref{the-spec} is the Chern-Simons interaction, with $\lambda_{\!A}$ controlling its strength and $\tilde{\bf{F}}^{\mu\nu}$ being the dual of $\bf{F}_{\mu\nu}$. The $S_{\rm{spec}}$ action is invariant under the local $SU(2)$ transformation
\Beq
\label{eq:gaugetransf1}
 {\bf{A}}_{\mu}&\rightarrow{\bf{U}}{\bf{A}_{\mu}}{\bf{U}}^{-1}+\frac{i}{g_{\!A}}(\nabla_{\mu}{\bf{U}}){\bf{U}}^{-1}\,,
\Eeq
where ${\bf{U}}=e^{-ig_{\!A}\boldsymbol{\beta}(x^{\nu})}$ with ${\boldsymbol{\beta}}=\beta^{a}\boldsymbol{\sigma}^a/2$. The action, $S_{\rm{matter}}$, is also invariant under the above local $SU(2)$ transformation, provided that the charged scalar transforms as $\boldsymbol{\varphi}\rightarrow{\bf{U}}\boldsymbol{\varphi}$.

In the (quasi) de Sitter geometry of slow-roll inflation
\Beq
\label{eq:dSst}
ds^2&=a^{2}(\tau)(d\tau^2-dx^2-dy^2-dz^2)\,,\\
a(\tau)&\simeq-\frac{1}{H\tau}\,,\quad \tau<0\,,\quad H\simeq\textrm{const}\,,
\Eeq
the gauge field value, $Q$, is a slowly varying quantity, and it corresponds to the following (almost) constant electric and magnetic fields
\be\label{Electric}
& E^a_{i} \equiv a E \delta^a_{i}\,, ~~&  \where E\simeq HQ(\tau)\,,\\
& B^a_{i} \equiv -a^2 B \delta^a_{i}\,, & \where B\simeq \ga Q^2(\tau)\,.
\ee 
In this setup, there is no preferred spatial direction. The equation of motion for the gauge field background is
\Beq
\label{eq:QEoM}
\partial_{\tau}^2Q+2\mathcal{H}\partial_{\tau}Q+(\partial_{\tau}\mathcal{H}+\mathcal{H}^2)Q+2a^2g_{\!A}^2Q^3=\frac{g_{\!A}\lambda_{\rm{A}}}{f}a\partial_{\tau}\chi Q^2-a^2\mathcal{J}\,,
\Eeq
where the spatially averaged component of the matter $3$-current is
\Beq
\label{eq:CurlyJ}
\mathcal{J}=\frac{1}{3a}\delta_{b}^{j}\langle J_{j}^{b}\rangle\,.
\Eeq
The covariantly conserved matter $4$-current, ${\bf{J}}_{\mu}=J^{a}_{\mu}\boldsymbol{\sigma}^a/2$, satisfying $\nabla^{\mu}{\bf{J}}_{\mu}=\boldsymbol{0}$, is given by the general expression
\Beq
\label{eq:Jdef}
J_{\mu}^{b}=\frac{\delta S_{\rm{matter}}}{\sqrt{-g}\delta A^{b\mu}}\,.
\Eeq
In our setup, the scalar doublet generates the matter $4$-current and its explicit form is
\Beq
\label{eq:CurrentHiggs}
J_{\mu}^a=\frac{ig_{\!A}}{2}\left[({{\bf{D}}}_{\mu}\boldsymbol{\varphi})^{\dagger}\boldsymbol{\sigma}^a\boldsymbol{\varphi}-\boldsymbol{\varphi}^{\dagger}\boldsymbol{\sigma}^a{{\bf{D}}}_{\mu}\boldsymbol{\varphi}\right]\,.
\Eeq

For our later convenience, we define the dimensionless quantity 
\be\label{xi_psi}
\xp\equiv\frac{g_{\!A}Q}{H},
\ee
which is equal to the ratio of the magnetic and electric fields, $\xp \simeq \frac{B}{E}$. In some papers, $\xp$ has been called $m_Q$, e.g., in \cite{Dimastrogiovanni:2016fuu,Agrawal:2017awz,Agrawal:2018mrg}. Avoiding instability in the scalar perturbations requires 
\be\label{xp-condition}
\xp>\sqrt{2},
\ee
 as was found for the original gauge-flation and chromo-natural models in Refs. \cite{Namba:2013kia} and \cite{Dimastrogiovanni:2012ew}, respectively. One mode in the scalar sector of these models would have a negative frequency at $\frac{k}{a} \propto (2-\xp^2)$ and avoiding instability requires the condition \eqref{xp-condition} (for more details see Refs. \cite{Adshead:2013nka} and \cite{Dimastrogiovanni:2012ew}). 


\section{Schwinger Effect by $SU(2)$ fields in de Sitter Universe}
\label{sec:SchwingerEffect}

In this section, we study the evolution of the charged scalar doublet in the presence of the $SU(2)$ gauge field and its corresponding Schwinger effect. In Fourier space, the field equation of $\boldsymbol{\varphi}$ takes the form
\Beq
\label{eq:EoMDoubletFS}
{\bf{L}}_{2\times2}({\boldsymbol{k}},\tau)\cdot\begin{pmatrix}\varphi_{\boldsymbol{k}}^1(\tau)\\ \varphi_{\boldsymbol{k}}^2(\tau)\end{pmatrix}=\boldsymbol{0}\,,
\Eeq
where, for the isotropic and homogeneous gauge field given in Eq. \eqref{eq:SU2Ansatz}, we have
\Beq
\label{eq:FullL}
{\bf{L}}_{2\times2}({\boldsymbol{k}},\tau)\equiv\left[\partial^2_{\tau}+2\mathcal{H}\partial_{\tau}+k^2+a^2(\tau)\left(\frac{3}{4}g_{\!A}^2Q^2(\tau)+m^2\right)\right]{\bf{I}}_{2\times2}-g_{\!A}a(\tau)Q(\tau)k^a\boldsymbol{\sigma}^a\,.
\Eeq
Since in general the last term in Eq. \eqref{eq:FullL} is non-diagonal, the two components of the scalar doublet, $\varphi^1$ and $\varphi^2$, are linearly coupled. However, we can easily diagonalize the term (and thereby diagonalize ${\bf{L}}_{2\times2}({\boldsymbol{k}},\tau)$). \footnote{Its time-independent eigenvectors can be used to make the columns of a unitary matrix
\Beq
{\bf{P}}({\boldsymbol{k}})=\begin{pmatrix*}[c]e^{-i\phi/2}\cos(\theta/2) & & e^{-i\phi/2}\sin(\theta/2)\\ e^{i\phi/2}\sin(\theta/2)\quad& &-e^{i\phi/2}\cos(\theta/2)\end{pmatrix*}\,,
\Eeq
where $\boldsymbol{k}=k\hat{\bk}$ and $\hat{\bk}=(\sin\theta\cos\phi,\sin\theta\sin\phi,\cos\theta)$. } In particular, we can expand the field in real space as 
\be
\Hgz(\tau,\bx) = \sum_{\lambda=\pm 1} \int d^3k \,e^{i\bk\cdot\bx}\hgz_{\lambda,\bk} {\bf e}^{\lambda}_{\bk}\,, 
\ee
where ${\bf e}^{\lambda}_{\bk}$ are the polarization doublets and are given by
\be
  {\bf e}^{+}_{\bk}=\left( \begin{matrix} e^{-i\frac{\phi}{2}} \cos\frac{\theta}{2} \\ e^{i\frac{\phi}{2}} \sin\frac{\theta}{2}  \end{matrix} \right)  \an  {\bf e}^{-}_{\bk}=\left( \begin{matrix}  e^{-i\frac{\phi}{2}} \sin\frac{\theta}{2} \\ -e^{i\frac{\phi}{2}} \cos\frac{\theta}{2}   \end{matrix} \right),
\ee
whereas the polar coordinates are defined by $\hat{\bk}=(\sin\theta\cos\phi,\sin\theta\sin\phi,\cos\theta)$. The polarization doublets satisfy the following relations
\be\label{e-sigma-e}
 {\bf e}^{\lambda\dag}_{\bk}\cdot{\bf e}^{\lambda'}_{\bk} = \delta_{\lambda\lambda'}\,,\qquad {\bf e}^{\lambda\dag}_{\bk}\cdot k^a\boldsymbol{\sigma}^a\cdot{\bf e}^{\lambda'}_{\bk} = \lambda \delta_{\lambda\lambda'} k\,.
\ee 
More explicitly, the new pair of fields are governed by the following decoupled equations of motion
\Beq
\label{eq:EoMPM}
\partial^2_{\tau}\varphi_{\lambda,\bk}+2\mathcal{H}\partial_{\tau}\varphi_{\lambda,\bk}+\left[k^2-\lambda a(\tau)g_{\!A}Q(\tau)k+a^2(\tau)\left(\frac{3}{4}g_{\!A}^2Q^2(\tau)+m^2\right)\right]\varphi_{\lambda,\bk}=0\,.
\Eeq
We now proceed to quantize the system by defining the canonically normalized fields
\Beq
\label{eq:qDefn}
q_{\lambda,\bk}=a(\tau)\varphi_{\lambda,\bk}\,.
\Eeq
The corresponding canonical conjugate momenta are
\Beq
\Pi_{\lambda,\bk}=\frac{\delta S_{\rm{spec}}}{\delta \partial_{\tau}q_{\lambda,\bk}}=\partial_{\tau}q_{\lambda,\bk}^*-\mathcal{H}q_{\lambda,\bk}^*\,.
\Eeq
We then promote $q_{\lambda,\bk}$ and their conjugate momenta to quantum operators, obeying the canonical commutation relations
\Beq
\left[q_{\lambda,\bk}(\tau),\Pi_{\lambda',\bk'}(\tau)\right]=i(2\pi)^{-3}\delta_{\lambda\lambda'}\delta^{3}(\bk-\bk')\,,
\Eeq
with all other commutators vanishing. In terms of creation and annihilation operators we have
\Beq
\label{eq:qpmGenSol}
q_{\lambda,\bk}=a_{\lambda,\bk}u_{\lambda,\bk}(\tau)+b_{\lambda,-\bk}^{\dagger}u_{\lambda,\bk}^*(\tau)\,,
\Eeq
where $a_{\lambda,\boldsymbol{k}}$ and $b_{\lambda,\boldsymbol{k}}$ are the annihilation operators of a particle with charge $+\ga$ and $-\ga$  with respect to the asymptotic past vacuum (see also Eq. \eqref{eq:VacDef}), respectively. If we normalize the mode functions as
\Beq
\label{eq:Wronskian}
u_{\lambda,\bk}\partial_{\tau}u_{\lambda,\bk}^*-u_{\lambda,\bk}^*\partial_{\tau}u_{\lambda,\bk}=i(2\pi)^{-3}\,,
\Eeq
then we have
\Beq
\left[a_{\lambda,\bk},a_{\lambda',\bk'}^{\dagger}\right]=\left[b_{\lambda,\bk},b_{\lambda,\bk'}^{\dagger}\right]=\delta_{\lambda\lambda'}\delta^3(\bk-\bk')\,,
\Eeq
with all other commutators of creation and annihilation operators vanishing.

The mode functions are governed by Eq. \eqref{eq:EoMPM} which can be reduced to
\Beq
\label{eq:EoMModeFunctions}
\partial_{\tau}^2u_{\lambda,\bk}+\omega_{\lambda,\bk}^2u_{\lambda,\bk}=0\,,
\Eeq
where 
\Beq
\label{eq:OmegaPM}
\omega_{\lambda,\bk}^2=k^2-\lambda ag_{\!A}Qk+a^2\left[\frac{3}{4}g_{\!A}^2Q^2+m^2-\frac{\partial_{\tau}^2a}{a^3}\right]\,.
\Eeq
Using Eq. \eqref{eq:dSst} and $Q=\rm{const}.$, as well as
\Beq\label{kappa-mu}
z\equiv2ki\tau\,,\qquad\kappa_{\lambda}\equiv-\lambda i\frac{\xp}{2}\,,\qquad\mu^2\equiv \frac{9}{4}-\frac{3}{4}\xp^2-\frac{m^2}{H^2}\,,
\Eeq
Eq. \eqref{eq:EoMModeFunctions} becomes
\Beq
\partial_{z}^2u_{\lambda,\bk}+\left[\frac{1}{z^2}\left(\frac{1}{4}-\mu^2\right)+\frac{\kappa_{\lambda}}{z}-\frac{1}{4}\right]u_{\lambda,\bk}=0\,,
\Eeq
with general solutions given by linear combinations of the Whittaker functions, $W_{\kappa_{\lambda},\mu}(z)$ and $M_{\kappa_{\lambda},\mu}(z)$. The solutions which reduce to the vacuum expressions at early times, i.e., when ${\rm Im}(z)\rightarrow-\infty$, are
\Beq
\label{eq:usExactSolution}
u_{\lambda,\bk}(z)=\frac{e^{i\kappa_{\lambda}\pi/2}}{(2\pi)^{3/2}\sqrt{2k}}W_{\kappa_{\lambda},\mu}(z)\,.
\Eeq
 One can explicitly check that the expression on the right hand side in Eq. \eqref{eq:usExactSolution} tends to 
\Beq
\label{eq:Minkowski}
\frac{e^{-ik\tau}}{(2\pi)^{3/2}\sqrt{2k}}\,,
\Eeq
in the limit of $k\tau\rightarrow-\infty$.  \footnote{The $W$ and $M$ Whittaker functions have the asymptotic expansions
\Beq\label{WM-asymp}
W_{\kappa,\mu} & \sim  e^{-z/2} z^{\kappa}\left[1+\mathcal{O}(z^{-1})\right] \qquad\quad &\textmd{for} \quad \lvert z\rvert \rightarrow \infty, \\
M_{\kappa,\mu} & \sim  z^{\mu+\frac12}\left[1+\mathcal{O}(z)\right] \quad &\textmd{for} \quad \lvert z\rvert \rightarrow 0,
\Eeq
implying that $W$/$M$ functions correspond to positive frequency modes in the asymptotic past/future limits, respectively (see Eq. \eqref{eq:AsympFutVacModeFunc}).}
Note that the effective frequency squared given in Eq. \eqref{eq:OmegaPM} can be negative for the $\lambda=+$ polarization state of the scalar field, e.g., for the interval $k|\Delta\tau|=2\sqrt{2-\frac{\xp^2}{2}}$ if \footnote{In this case, $\omega_{\lambda=+,\bk}^2=0$ has two real roots, $k\tau_{1,2}=-\xp/2\pm\sqrt{2-\xp^2/2}\,$.} $\xp<2$ and $m=0$. Therefore, when $\xp<2$, the scalar field with the $\lambda=+$ polarization state can experience a small enhancement in comparison to the $\lambda=-$ state around the horizon crossing. However, the gauge field generates an extra mass for the scalar doublet as
\be
m_{\rm eff}^2 = m^2 + \frac34 \xp^2 H^2.
\ee 
The super-horizon behavior of the scalar field amplitude goes as 
$\varphi_{\lambda,\bk} \propto a^{-\frac32+\sqrt{\frac94- \frac{m_{\rm eff}^2}{H^2}}}$. Given that $m^2_{\rm{eff}}/H^2\geq3\xp^2/4$ and $\xp>\sqrt{2}$, it implies that $\varphi_{\lambda,\bk}$ always decays faster than $a^{-\frac35}$. Therefore, the scalar field always damps after horizon crossing and does not contribute to the super-horizon scalar perturbations in the inflation sector. 


\subsection{Schwinger pair production}

In Section \ref{sec:SchwingerEffect} we showed that the scalar field modes coupled to the $SU(2)$ gauge field background in de Sitter spacetime get excited. The mode functions change from the early-time ($k\tau\rightarrow-\infty$) form given in Eq. \eqref{eq:Minkowski} to the late-time excited form given in Eq. \eqref{eq:usExactSolution}. In order to interpret these scalar field excitations as particles, we need to be in the adiabatic limit
\Beq
\label{eq:AdiabCond}
\left(\frac{\partial_{\tau}\omega_{\lambda,\boldsymbol{k}}}{\omega^{2}_{\lambda,\boldsymbol{k}}}\right)^2\ll1\, \an\left|\frac{\partial_{\tau}^2\omega_{\lambda,\boldsymbol{k}}}{\omega^{3}_{\lambda,\boldsymbol{k}}}\right|\ll1\,.
\Eeq
Only then we can have a well-defined late-time adiabatic vacuum. The field excitations about it play the role of particles. 

In (quasi) de Sitter spacetime with $Q\simeq\rm{const}.$, the frequencies given in Eq. \eqref{eq:OmegaPM} can be written as
\Beq
\label{eq:OmegaPMdS}
\omega_{\lambda,\boldsymbol{k}}^{2} \approx\begin{cases}
    k^2,& \text{if } k\tau\rightarrow-\infty\\\
    \dfrac{1}{\tau^2}\left(\dfrac{3\xp^2}{4}+\dfrac{m^2}{H^2}-2\right),              & \text{if } k\tau\rightarrow0\,.
\end{cases}
\Eeq
At early times, $k\tau\rightarrow-\infty$, the adiabaticity conditions given in Eq. \eqref{eq:AdiabCond} are trivially satisfied and the adiabatic vacuum is the Bunch-Davies vacuum \cite{Birrell:1982ix}. That is what we take as an initial condition, see Eq. \eqref{eq:Minkowski}. At later times the adiabaticity conditions can be violated. However, in the asymptotic future, $k\tau\rightarrow0$,
\Beq
\left(\frac{\partial_{\tau}\omega_{\lambda,\boldsymbol{k}}}{\omega^{2}_{\lambda,\boldsymbol{k}}}\right)^2\approx\left(\frac{3\xp^2}{4}+\frac{m^2}{H^2}-2\right)^{-1}\, \an \frac{\partial_{\tau}^2\omega_{\lambda,\boldsymbol{k}}}{\omega^{3}_{\lambda,\boldsymbol{k}}}\approx2\left(\frac{3\xp^2}{4}+\frac{m^2}{H^2}-2\right)^{-1}\,,
\Eeq
they can become again adiabatic if
\Beq
\label{eq:AdiabCond2}
\frac{3\xp^2}{4}+\frac{m^2}{H^2}\gg1\,.
\Eeq
From the combination of the above conditions, we find that $\mu$ in Eq. \eqref{kappa-mu} should be a pure imaginary quantity and for the rest of this section, we write
\Beq
\mu=i \lvert \mu \rvert,
\Eeq 
and assume that the condition given in Eq. \eqref{eq:AdiabCond2} holds. Then the new asymptotic future vacuum mode functions of $q_{\lambda,\boldsymbol{k}}$ are (see Eq. \eqref{WM-asymp})
\Beq
\label{eq:AsympFutVacModeFunc}
v_{\lambda,\boldsymbol{k}}(-2i\x)=\frac{e^{-|\mu|\pi/2}}{(2\pi)^{3/2}2\sqrt{k|\mu|}}M_{\kl,\mu}(-2i\x)\,.
\Eeq
They agree up to an unphysical constant phase with the positive-frequency WKB solutions of Eq. \eqref{eq:EoMModeFunctions} in the asymptotic future, $k\tau\rightarrow0$,
\Beq
\frac{e^{-i\int^{\tau}\omega^{\pm}_{\boldsymbol{k}}d\tau}}{(2\pi)^{3/2}\sqrt{2\omega^{\pm}_{\boldsymbol{k}}}}\rightarrow\frac{(-\tau)^{i|\mu|+1/2}e^{i\cdot\rm{const}}}{(2\pi)^{3/2}\sqrt{2|\mu|}}\,.
\Eeq

By expanding $q_{\lambda,\boldsymbol{k}}$ in terms of the annihilation and creation operations of the asymptotic future vacuum we write
\Beq\label{u-2}
q_{\lambda,\boldsymbol{k}}=(\tilde{a}_{\lambda,\boldsymbol{k}} v_{\lambda,\boldsymbol{k}} + \tilde{b}^{\dag}_{\lambda,-\boldsymbol{k}} v^{*}_{\lambda,-\boldsymbol{k}} \big)\,.
\Eeq
Then we have
\Beq
\label{eq:VacDef}
a_{\lambda,\boldsymbol{k}} \lvert 0_{\rm in}\rangle =0 &\an& b_{\lambda,\boldsymbol{k}}\lvert 0_{\rm in} \rangle =0,\\
\tilde{a}_{\lambda,\boldsymbol{k}}\lvert 0_{\rm out}\rangle =0 &\an& \tilde{b}_{\lambda,\boldsymbol{k}}\lvert 0_{\rm out} \rangle =0,
\Eeq
where $\lvert 0_{\rm in}\rangle$ and $\lvert 0_{\rm out}\rangle$ are the vacuum states in the asymptotic past, $k\tau\rightarrow-\infty$, and future, $k\tau\rightarrow0$, of the (quasi) de Sitter spacetime, respectively.
Using Bogoliubov transformations, we can relate $\tilde{a}_{\lambda,\boldsymbol{k}}$ and $\tilde{b}_{\lambda,\boldsymbol{k}}$ to $a_{\lambda,\boldsymbol{k}}$ and $b_{\lambda,\boldsymbol{k}}$ as 
\Beq
\tilde{a}_{\lambda,\boldsymbol{k}}&= \alpha_{\lambda,\boldsymbol{k}} a_{\lambda,\boldsymbol{k}} + \beta^{*}_{\lambda,\boldsymbol{k}} b^{\dag}_{\lambda,-\boldsymbol{k}},\\
\tilde{b}_{\lambda,\boldsymbol{k}}&= \alpha_{\lambda,-\boldsymbol{k}} b_{\lambda,\boldsymbol{k}} + \beta^{*}_{\lambda,-\boldsymbol{k}} a^{\dag}_{\lambda,-\boldsymbol{k}},
\Eeq
in which $\alpha_{\lambda,\boldsymbol{k}}$ and $\beta_{\lambda,\boldsymbol{k}}$ are four Bogoliubov coefficients satisfying the normalization condition
\Beq
|\alpha_{\lambda,\boldsymbol{k}}|^2-|\beta_{\lambda,\boldsymbol{k}}|^2=1\,.
\Eeq
Comparing Eqs. \eqref{eq:usExactSolution} and \eqref{eq:AsympFutVacModeFunc}, we find \footnote{When reading the Bogoliubov coefficients, we used the following relation between the $W$ and $M$ functions (which holds when $2\mu$ is not an integer and $ -\frac{3\pi}{2}< \lvert \rm{arg}z\rvert <\frac{\pi}{2}$) \cite{Sp-book,Nist}
\Beq
M_{\kappa,\mu}(z)= \frac{\Gamma(2\mu+1)}{\Gamma(\frac12+\mu+\kappa)} e^{-i(\frac12+\mu-\kappa)\pi} W_{\kappa,\mu}(z) + \frac{\Gamma(2\mu+1)}{\Gamma(\frac12+\mu-\kappa)} e^{i\kappa \pi} W_{-\kappa,\mu}(e^{i\pi}z).
\Eeq
Moreover, using the relation $W_{\kappa,-\mu}(z)=W_{\kappa,\mu}(z)$ and the fact that in our case $z$, $\mu$ and $\kappa$ are all imaginary quantities, we have $W^{*}_{\kappa,\mu}(z)=W_{-\kappa,\mu}(-z)$.  }
\Beq\label{alpha-beta}
\alpha_{\lambda,\boldsymbol{k}} &= \sqrt{2\lvert \mu\rvert } e^{(\lambda \lvert\kl\rvert+\lvert \mu\rvert)\pi/2} \frac{\Gamma(-2\mu)}{\Gamma(\frac12-\mu-\kl)}  , \\
\beta_{\lambda,\boldsymbol{k}} &= -i \sqrt{2\lvert \mu\rvert } e^{(\lambda\lvert\kl\rvert-\lvert \mu\rvert)\pi/2} \frac{\Gamma(2\mu)}{\Gamma(\frac12+\mu-\kl)}  .
\Eeq
Note that in the adiabatic limit these Bogoliubov coefficients become independent of the momentum and (almost) constant in a (quasi) de Sitter spacetime
\Beq
\alpha_{\lambda,\boldsymbol{k}} = \alpha_{\lambda} \an \beta_{\lambda,\boldsymbol{k}}= \beta_{\lambda}.
\Eeq
Of course, the full solution of $\beta_{\lambda,\bk}$ must vanish in the $k\rightarrow \infty$ limit.


\subsection{Pair production rate and vacuum-vacuum transition amplitude}

Having the $\beta_{\lambda,\boldsymbol{k}}$ coefficient, we are ready to determine the particle number density as well as the vacuum-vacuum transition amplitude. The number density of the created particles with charge $\pm \ga$ and a given comoving momentum, $\bk$, in the asymptotic future is
\Beq
n_{\lambda}(\bk) = \langle 0_{\rm in}\rvert \tilde{a}^{\dag}_{\lambda,\boldsymbol{k}}\tilde{a}_{\lambda,\boldsymbol{k}} \lvert 0_{\rm in}\rangle   = \langle 0_{\rm in}\rvert \tilde{b}^{\dag}_{\lambda,-\boldsymbol{k}}\tilde{b}_{\lambda,-\boldsymbol{k}} \lvert 0_{\rm in}\rangle = \lvert \beta_{\lambda} \rvert^2= \frac{e^{2\lambda \lvert \kappa\rvert \pi}+e^{-2\lvert\mu\rvert\pi}}{2\sinh(2\lvert\mu\rvert\pi)},
\Eeq
which has a $k$-independent spectrum for each polarization state in the regime where the solution is valid. The total particle creation from the asymptotic past to the asymptotic future, therefore, is
\be
N_{\lambda} = \frac{1}{(2\pi)^3}  \lvert \beta_{\lambda} \rvert^2 \int d^3k = \frac{1}{(2\pi)^2}  \frac{e^{\lambda \xp \pi}+e^{-2\lvert\mu\rvert\pi}}{\sinh(2\lvert\mu\rvert\pi)} \int^{\infty}_{0} k^2dk,
\ee 
which appears to be divergent. This makes physical sense since it expresses the number of pairs created for all times. A more physically meaningful quantity is the pair production rate, i.e., the number of pairs produced per unit time per unit physical volume
\Beq
\label{eq:DecayDef}
\Gamma^{\lambda}_{\rm{pairs}}=\frac{1}{a(\tau)^{4}}\frac{dN_{\lambda}}{d\tau}\,.
\Eeq
To calculate the derivative we need to convert the wavenumber integral into a time integral. We know that for $\lvert \mu \rvert^2\gg1$, the asymptotic past and future are adiabatic vacua for the field and therefore no particle production occurs in the infinite past and future. A pair of particles of a given comoving momentum, $k$, is produced only at a single non-adiabatic event, when the adiabaticity conditions, Eq. \eqref{eq:AdiabCond}, are violated. Given $|\mu|^2\gg1$, an approximate estimate for the moment when this happens is
\Beq\label{tau-k}
\tau(k)\approx -\frac{1}{k}\left(\frac{3\xp^2}{4}+\frac{m^2}{H^2}-2\right)^{\frac12} \approx -\frac{1}{k}|\mu|\,.
\Eeq
 Hence,
\Beq
N_{\lambda}\approx |\mu|^3\frac{e^{\lambda \xp\pi}+e^{-2|\mu|\pi}}{(2\pi)^2\sinh(2|\mu|\pi)}\int_{-\infty}^0 d\tau \,(a(\tau)H)^4\,,
\Eeq
yielding constant pair production rates
\Beq
\Gamma^{\lambda}_{\rm{pairs}}\approx |\mu|^3\frac{e^{\lambda \xp\pi}+e^{-2|\mu|\pi}}{(2\pi)^2\sinh(2|\mu|\pi)}H^4\,.
\Eeq
Integrating Eq. \eqref{eq:DecayDef}, we find that the physical number densities of pairs created up to a time $\tau$ are also time independent
\Beq
\label{eq:nPairs}
n_{\rm{pairs}}^{\lambda}=\frac{1}{a(\tau)^3}\int_{-\infty}^{\tau'} d\tau\, a(\eta)^4 \Gamma_{\rm{pairs}}^{\lambda}\approx\frac{\Gamma_{\rm{pairs}}^{\lambda}}{3H}\,,
\Eeq
i.e., gravitational and Schwinger particle production are exactly balanced by the the gravitational redshifting, in the limit given in Eq. \eqref{eq:AdiabCond2}. 

The production rate of the pairs with $\lambda=\pm$ polarization states are different and the $\lambda=+$ state is generated much more. In particular, in the  $\lvert \mu\rvert^2\gg 1$ limit, we have
\be
\frac{\Gamma^{-}_{\rm{pairs}}}{\Gamma^{+}_{\rm{pairs}}} \approx e^{-2\xp \pi}.
\ee
Finally, summing over the two polarization states of the scalar field, we have 
\Beq
\label{eq:nPairs-II}
n_{\rm{pairs}}= \frac{2H^3}{3(2\pi)^2} |\mu|^3\frac{\cosh( \xp \pi)+e^{-2|\mu|\pi}}{\sinh(2|\mu|\pi)}\,.
\Eeq
Since $\lvert \mu\rvert^2\gg 1$, we can approximate the above as 
\Beq
\label{eq:nPairs-III}
n_{\rm{pairs}}\approx \frac{2H^3}{3(2\pi)^2} |\mu|^3 e^{(\xp-2|\mu|) \pi}\,,
\Eeq
which is exponentially suppressed because $2|\mu|-\xp\gg1$. That is due to the fact that $\beta_{\lambda}$ coefficients in Eq. \eqref{alpha-beta} have always a negative exponential power, i.e.,
$$i\kappa_{\lambda} - \lvert \mu \rvert <0,$$
regardless of the parameters. 

The vacuum-vacuum transition can be computed as
\be
\lvert \langle 0_{\rm out}\vert 0_{\rm in} \rangle \rvert ^2 \equiv e^{(-\iint d^3xd\tau a^4 \Upsilon_{\rm vac})} = \exp \bigg[ - \frac{1}{(2\pi)^3}\int d^3x  \int d^3k \ln(1+ \lvert \beta_{\lambda,\boldsymbol{k}}\rvert^2) \bigg]\,,
\ee 
where $\Upsilon_{\rm vac}$ is the vacuum decay rate. Using Eq. \eqref{tau-k}, we obtain
\be
 \Upsilon_{\rm vac} =  - 2 \sum_{\lambda=\pm1} \frac{H^4}{(2\pi)^2} \lvert \mu \rvert^3 \ln \bigg[ 1+  \bigg( \frac{1+e^{-2(\lvert \mu \rvert-\lambda \lvert \kappa \rvert )\pi} }{\sinh(2\lvert \mu \rvert \pi)} \bigg)\bigg]\,.
\ee
Since $\lvert \mu \rvert^2\gg1$, the decay rate is well approximated by
\be\label{vacuum-decay}
 \Upsilon_{\rm vac} \approx  -  \frac{H^4}{\pi^2} \lvert \mu \rvert^3 e^{-2\lvert \mu \rvert \pi}\,,
\ee
yielding an exponential suppression.


\subsection{Minkowski limit}

Let us consider the Minkowski limit. Using the electric field (see Eq. \eqref{Electric})
\be
\xp = \frac{\ga E}{H^2},
\ee
in Eqs. \eqref{eq:nPairs} and \eqref{vacuum-decay} and taking the $H\rightarrow 0$ limit, while keeping $E=\rm{const}$, we have
\be
\Gamma_{\rm{pairs}}^{\lambda} \rightarrow 0 \an \Upsilon_{\rm vac} \rightarrow0.
\ee
The Bogoliubov coefficients in this limit are 
\be
\alpha_{\lambda,\bk}= 1 \an \beta_{\lambda,\bk} =0.
\ee 
Therefore, our setup with an isotropic $SU(2)$ gauge field does not experience the Schwinger effect in the Minkowski limit and the definition of vacuum remains unchanged with time. 
Notice that it is unlike the standard constant $U(1)$ electric field in flat space in which both $\Gamma_{\rm{pairs}}^{\lambda}$ and $\Upsilon_{\rm vac}$ are finite and constant \cite{Kobayashi:2014zza}. 
 The main difference between a $U(1)$ gauge field in the direction $\hat{z}$, and an isotropic $SU(2)$ gauge field lies in the values of their $\lvert \mu\rvert$ and $\lvert \kappa\rvert$. More precisely, in the massless limits, for a $U(1)$ constant electric field, modes with momentum $k=k_z$ have $\lvert \mu\rvert^2\simeq \lvert \kappa\rvert^2\sim (\frac{eE}{H^2})^2$, while in the isotropic $SU(2)$ gauge field case, we have $\lvert \mu\rvert^2\simeq 3\lvert \kappa\rvert^2\sim \frac34(\frac{g_AE}{H^2})^2$. Therefore, in the latter case, $\Gamma_{\rm{pairs}}^{\lambda}$ and $\Upsilon_{\rm vac}$ are exponentially suppressed.


\section{The Induced Current by an $SU(2)$ gauge field}
\label{sec:Indcurr}

In the previous section, we showed that the individual comoving modes can be amplified both gravitationally and due to the existence of non-Abelian background fields. We now calculate the induced current, $J_{\mu}^a$, given in Eq. \eqref{eq:CurrentHiggs}, as a consequence of this amplification.

Let us take the expectation value of $J_{\mu}^a$ with respect to the vacuum state of the asymptotic past, $  \lvert 0_{\rm in}\rangle$. Writing the Fourier transforms of $\boldsymbol{\varphi}$ in terms of $q_{\lambda,\bk}$, we find that the expectation value of the charge density, $J_0^a$, for the background configuration in Eq. \eqref{eq:SU2Ansatz}, vanishes
\Beq
 \langle 0_{\rm in}\lvert J_{0}^a \rvert  0_{\rm in}\rangle =0\,.
\Eeq
In deriving this result, we used the relations in Eq. \eqref{e-sigma-e}. The above result makes physical sense since particles are created in pairs and the net charge density should be zero.

The spatial components are
\Beq
\label{eq:Jaj}
\langle0_{\rm in}|J_{j}^a|0_{\rm in}\rangle=-\frac{g_{\!A}}{a^2}\sum_{\lambda=\pm 1}\int d^3k\left(\lambda\frac{k^jk^a}{k}-\delta^{a}_j\frac{g_{\!A}}{2}aQ\right)|u_{\lambda,\bk}|^2\,.
\Eeq
We used the identity $\{\boldsymbol{\sigma}^a,\boldsymbol{\sigma}^b\}=2\delta^{ab}{\bf{I}}_{2\times2}$ as well as Eq. \eqref{e-sigma-e}.
Using Eq. \eqref{eq:usExactSolution}, Eq. \eqref{eq:Jaj} can be further reduced to
\Beq
\label{eq:JJ}
\langle0_{\rm in}|J_{j}^a|0_{\rm in}\rangle=\delta^{a}_ja(\tau)\mathcal{J}\,,
\Eeq
where $\mathcal{J}$ is defined by Eq. \eqref{eq:CurlyJ} and is given by
\Beq\label{curly-J}
\mathcal{J}=\frac{g_{\!A}}{(2\pi)^2a^3}  \lim_{\Lambda\rightarrow \infty} \int^{aH \Lambda} k^3d\ln k\sum_{\lambda=\pm1}\left(-\lambda\frac{k}{3}+\frac{g_{\!A}}{2}aQ\right)\frac{e^{i\kappa_{\lambda}\pi}}{k}|W_{\kappa_{\lambda},\mu}(2ki\tau)|^2\,.
\Eeq
In the above, $\Lambda$ is the UV cutoff on the physical momentum, $\Lambda= \frac{k_{\rm{UV}}}{aH}$, which we eventually send to infinity. 
We will later also calculate the conductivity defined by
\be\label{conductive}
\sigma \equiv \frac{\mathcal{J}}{E} = \frac{1}{\xp} \frac{\ga \mathcal{J}}{H^2}.
\ee

We work out the explicit form of $\mathcal{J}$ in Appendix \ref{Current-appendix}. Here we only report the final result.
From the combination of Eqs. \eqref{cJ-to-cG} and \eqref{G-tot}, we have 
\be\label{cJ-tot}
\mathcal{J}(\Lambda) = &-& \lim_{\Lambda\rightarrow \infty} \frac{\ga H^3}{(2\pi)^2}
 \bigg\{  - \frac13 \xp  \Lambda^{2} - \frac{1}{6} \xp \big(1+ \xp^2 \big) \ln(2\Lambda)   +\frac{1}{6} \xp \left(\frac{25}{12} - \mu^2 + \frac{13}{12} \xp^2\right)  \nonumber\\
 &+&  \frac{1}{24} \xp(1+\xp^2) ~{\rm{Re}} \bigg[ \frac{(e^{\xp\pi}+e^{2\lvert\mu\rvert\pi})}{\sinh(2\lvert\mu\rvert\pi)} \psi\left(\frac12+\frac{i}{2}\xp-\mu\right) \nonumber\\
&-&   \frac{(e^{\xp\pi}+e^{-2\lvert\mu\rvert\pi})}{\sinh(2\lvert\mu\rvert\pi)} \psi\left(\frac12+\frac{i}{2}\xp+\mu\right)+   \frac{(e^{-\xp\pi}+e^{2\lvert\mu\rvert\pi})}{\sinh(2\lvert\mu\rvert\pi)} \psi\left(\frac12-\frac{i}{2}\xp-\mu\right) \nonumber\\
&-&   \frac{(e^{-\xp\pi}+e^{-2\lvert\mu\rvert\pi})}{\sinh(2\lvert\mu\rvert\pi)} \psi\left(\frac12-\frac{i}{2}\xp+\mu\right) \bigg] +  \frac{\mu}{9}(4-4\mu^2 +3\xp^2)\frac{\sinh(\xp \pi)}{\sin(2\mu\pi)} 
  \bigg\},
\ee
where $\psi(z)\equiv\partial_z\Gamma(z)/\Gamma(z)$ and $\Gamma(z)$ is the gamma function. $\mathcal{J}(\Lambda)$ has a quadratic and a logarithmic UV divergent terms. To deal with this divergent behaviour, we follow \cite{Kobayashi:2014zza} and regularize the current using the method of adiabatic subtraction. We define the regularized current as
\Beq
\langle 0_{\rm in}|J_{j}^a| 0_{\rm in}\rangle_{\rm{reg}}=\langle 0_{\rm in}|J_{j}^a|0_{\rm in} \rangle-\langle 0_{\rm in}|J_{j}^a| 0_{\rm in} \rangle_{{\rm WKB}}^{(2)}\,,
\Eeq
where the last term is the expectation value of the current density given in Eq. \eqref{eq:Jaj} with $u^{\pm}_{\boldsymbol{k}}\rightarrow u^{\pm}_{\boldsymbol{k}}|^{(2)}_{\rm WKB}$ and $u^{\pm}_{\boldsymbol{k}}|^{(2)}_{\rm WKB}$ are the WKB solutions up to second adiabatic order. We present the details of computations in Appendix \ref{Adiabatic Subtraction} and here we only report the final result: 
\be\label{cJ-WKB}
\mathcal{J}^{(2)}_{\rm{WKB}}(\Lambda) = &-& \lim_{\Lambda\rightarrow \infty} \frac{\ga H^3}{(2\pi)^2} 
 \bigg[ - \frac13 \xp  \Lambda^{2}   - \frac{1}{6} \xp \big(1+ \xp^2 \big) \ln\Lambda  +\frac{\xp}{12}(1+\xp^2) \ln\left(\frac{9-4\mu^2 -\xp^2}{4}\right) \nonumber\\
&+& \frac{\xp}{6}\left(\frac{25}{12} - \mu^2 + \frac{13}{12}\xp^2 \right)   +  \frac{\xp^3}{9} \frac{(\frac{13}{2}\xp^2 + 28 \mu^2 - 63)}{(-9+4\mu^2+\xp^2)^2} \bigg].
\ee
Subtracting Eq. \eqref{cJ-WKB} from Eq. \eqref{cJ-tot}, we find the desired regularized current
\be\label{cJ-reg}
  \mathcal{J}_{\rm{reg}} &=&  \frac{\ga H^3}{(2\pi)^2} 
 \bigg\{ 
 \frac{\xp^3}{9} \frac{(\frac{13}{2}\xp^2 + 28 \mu^2 - 63)}{(-9+4\mu^2+\xp^2)^2} + \frac{\xp}{12}(1+\xp^2) \ln\left(\frac{9-4\mu^2 -\xp^2}{4}\right)  \nonumber\\
&-&  \frac{4\mu}{9}\left(1-\mu^2 +\frac34\xp^2\right)\frac{\sinh\xp \pi}{\sin2\mu\pi} -  \frac{1}{24} \xp(1+\xp^2) ~{\rm{Re}} \bigg[ \frac{(e^{\xp\pi}+e^{2\lvert\mu\rvert\pi})}{\sinh(2\lvert\mu\rvert\pi)} \psi\left(\frac12+\frac{i}{2}\xp-\mu\right)\nonumber\\
  &-&   \frac{(e^{\xp\pi}+e^{-2\lvert\mu\rvert\pi})}{\sinh(2\lvert\mu\rvert\pi)} \psi\left(\frac12+\frac{i}{2}\xp+\mu\right) +   \frac{(e^{-\xp\pi}+e^{2\lvert\mu\rvert\pi})}{\sinh(2\lvert\mu\rvert\pi)} \psi\left(\frac12-\frac{i}{2}\xp-\mu\right) \nonumber\\
  & - &  \frac{(e^{-\xp\pi}+e^{-2\lvert\mu\rvert\pi})}{\sinh(2\lvert\mu\rvert\pi)} \psi\left(\frac12-\frac{i}{2}\xp+\mu\right) \bigg]  \bigg\}\,.
\ee
The conductivity is then given by Eq. \eqref{conductive} for $\mathcal{J}_{\rm reg}$. 


\subsection{Results}

\begin{figure*}[t] 
   \centering
  \includegraphics[width=3in]{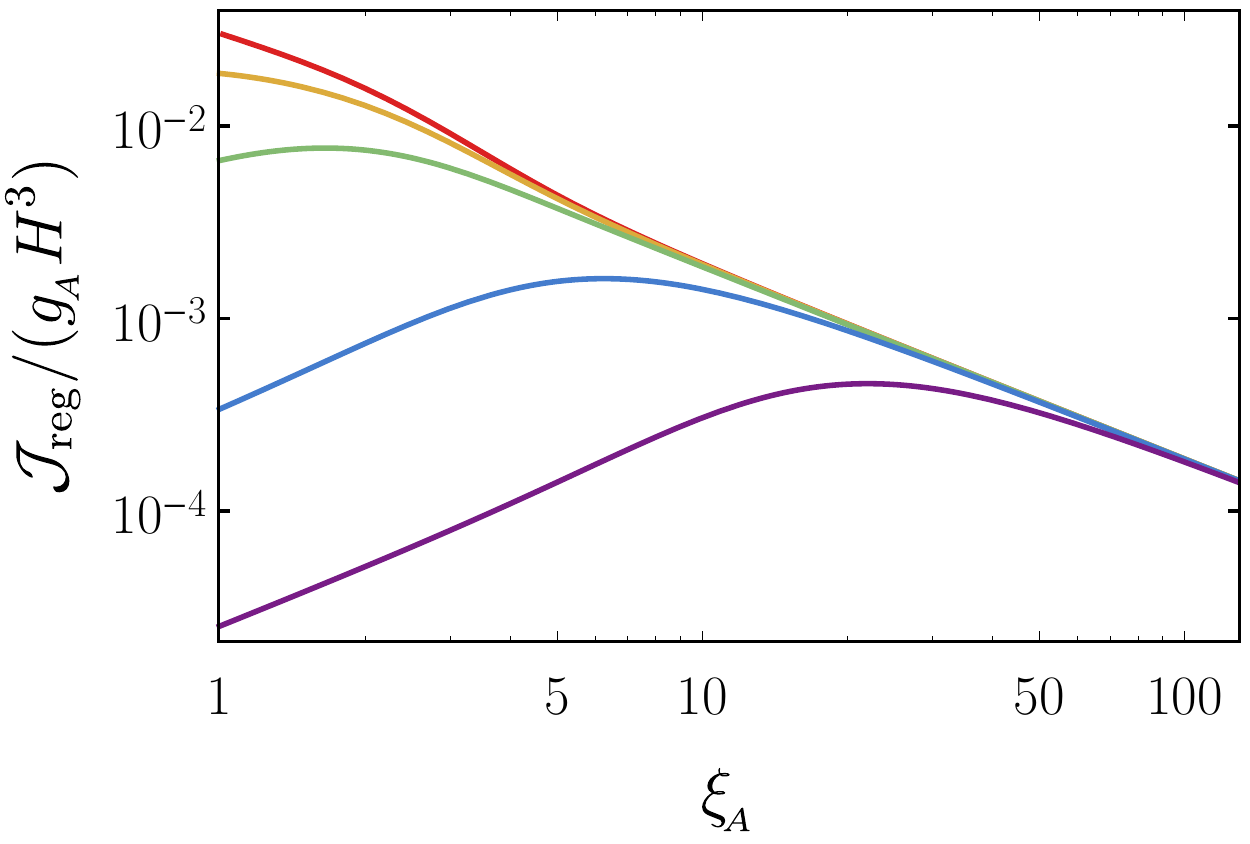} 
  \hspace{0.05in}
  \raisebox{0.295\height}{\includegraphics[height=1.555in]{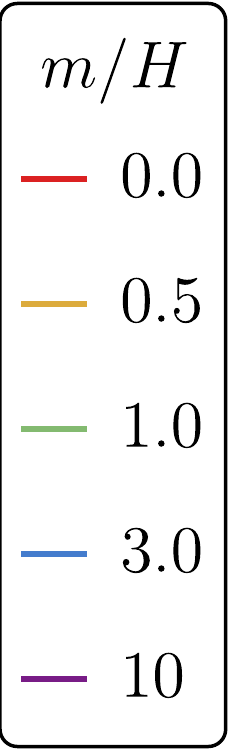}}\\
  \vspace{0.5in}
  \includegraphics[width=3in]{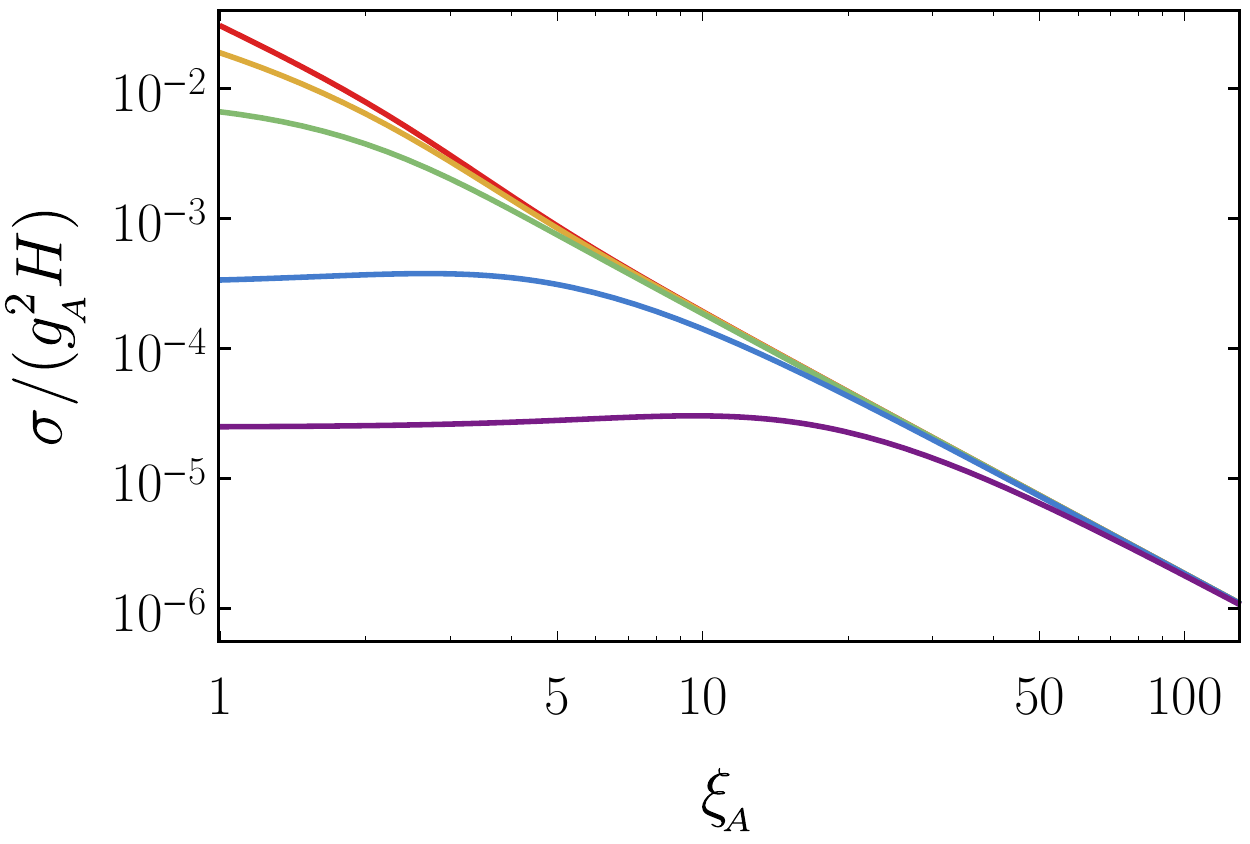}
  \hspace{0.05in}
  \raisebox{0.295\height}{\includegraphics[height=1.555in]{LegJmQ.pdf}}
   \caption{The induced current and conductivity of the scalar doublet field as a function of $\xp$ and $m/H$. The ratio of $\frac{\xp}{m/H}$ divides the system into two different regimes. For $\xp \lesssim 2 \frac{m}{H}$, the current is increasing linearly with $\xp$ and the conductivity is nearly independent of $\xp$. For $\xp \gtrsim 2 \frac{m}{H}$, the current is decreasing like $\xp^{-1}$. Both the current and conductivity decrease monotonically as $m/H$ increases.}
   \label{fig:JmQ}   
\end{figure*}

In Fig. \ref{fig:JmQ}, we show the current and the conductivity as a function of $\xp$ for different values of the scalar doublet mass, $m/H$. The current has some interesting features which we summarize in the following.

\begin{itemize}
\item{ The dimensionless quantity $\mathcal{J}_{\rm{reg}}/(\ga H^3)$ depends only on $\xp$ and $\frac{m}{H}$.}
\item{The current is an odd function of $\xp$; thus, $\mathcal{J}_{\rm{reg}}(\xp=0)=0$. As a result, the conductivity is an even function of $\xp$. Besides, for all positive (negative) values of $\xp$, the current is positive (negative).}
\item{The current increases (decreases) with $\xp$ in the limits that $\xp\lesssim \frac{m}{H}$ ($\xp\gg 1$ and $\xp\gtrsim \frac{m}{H}$). Thus, for any given value of $\frac{m}{H}$, $\mathcal{J}_{\rm reg}$ has an absolute maximum, $\mathcal{J}^{\rm Max}_{\rm reg} \sim \frac{7}{20} \frac{\ga H^3}{(2\pi)^2}\frac{1}{2m}$, at $\xp\sim 2 \frac{m}{H}$. In the following we focus on these two regimes and the corresponding asymptotic behaviors of the induced current and conductivity.}
\end{itemize} 
 

\subsubsection{Small $\xp$ limit: $\xp\lesssim \frac{m}{H}$}

As mentioned in Section \ref{sec:setup}, our setup has a lower bound on $\xp$ as $\xp\geq \sqrt{2}$. Therefore, by small $\xp$ limit we really mean $\sqrt{2}\leq\xp \lesssim \frac{m}{H}$ with $\frac{m}{H} \gtrsim 2$. 

In this limit, the current increases linearly with $\xp$ and has the form 
\be
\mathcal{J}_{\rm{reg}}(\xp,m) \sim \frac{7/20}{2(2\pi)^2} \frac{\xp}{m/H} \ga H^3,
\ee
which gives a $\xp$-independent conductivity, i.e., $\sigma \sim \frac{7/20}{2(2\pi)^2} \frac{\ga^2 H}{m/H}$. In other words, in the small $\xp$ limit, the conductivity is independent of $\xp$ while both the current and conductivity decrease like $(\frac{m}{H})^{-1}$ as the mass increases.


\subsubsection{Large $\xp$ limit: $\xp\gg 1$ and $\xp\gtrsim \frac{m}{H}$}

Here we turn to the large values of $\xp$, $\xp \gtrsim \frac{m}{H}$, and find the asymptotic form of the current. In deriving the explicit form of $\mathcal{J}_{\rm{reg}}$, we take the limit with $\xp\gg \frac{m}{H}$. Using the asymptotic expansion of the digamma functions \footnote{Here we used the asymptotic series for the gamma function
\be
\lim_{\lvert z \rvert\rightarrow \infty}\Gamma(z)\simeq \sqrt{2\pi} z^{z-\frac12}\exp(-z) \bigg(1+\frac{1}{12z}+\frac{1}{288z^2}-\frac{139}{51840z^3} + \dots \bigg). \nonumber
\ee} 
\be
\psi(z) = \ln(z)-\frac{1}{2z}-\frac{1}{12z^2}+\frac{1}{120z^4} + \dots\,\,\,, \where \lvert z \rvert \rightarrow \infty\,,
\ee
in Eq. \eqref{cJ-tot} with  $z_{\pm}= \frac12 - \mu \pm i \frac12\xp$ and $\mu \simeq \frac{i\sqrt{3}}{2}\xp \big(1-\frac{3}{2\xp^2} \big)+\mathcal{O}(\xp^{-3})$,
we find the finite terms in $\mathcal{J}$ as
\be\label{cJ-xi-L}
\mathcal{J} \supset -\frac{1}{6} \frac{\ga H^3}{(2\pi)^2}
 \bigg[  \xp \left(\frac{25}{12} - \mu^2 + \frac{13}{12} \xp^2\right) + \xp(1+\xp^2)  \ln\left(\frac{\xp}{\sqrt{2}}\right) - \frac{1}{12} \left(29 \xp + \frac{401}{5\xp}\right) \bigg]\,, \nonumber\\
\ee 
while $\mathcal{J}^{(2)}_{\rm{WKB}}$ given in Eq. \eqref{cJ-WKB} has the following finite terms
\be
\mathcal{J}^{(2)}_{\rm{WKB}} \supset -\frac{1}{6} \frac{\ga H^3}{(2\pi)^2} \bigg[ \xp(1+\xp^2) \ln\left(\frac{\xp}{\sqrt{2}}\right)  + \xp \left(\frac{25}{12} - \mu^2 + \frac{13}{12}\xp^2 \right) - \frac{29}{12} \xp\bigg] .
\ee
As a result $\mathcal{J}_{\rm reg}$ in the $\xp\gg m/H$ limit is
\be
\mathcal{J}_{\rm reg}\simeq  \frac{401}{360} \frac{\ga H^3}{(2\pi)^2} \xp^{-1}.
\ee
In the limit that $\xp\gg 1$ and $\xp \gg \frac{m}{H}$, by increasing the $\xp=\frac{B}{E}$ value, the current decreases like $1/\xp$ and the conductivity like $1/\xp^2$.


\subsection{Comparison with the Schwinger effect by a $U(1)$ field in de Sitter space}

In this section we compare the particle production and induced current in our setup with those by a $U(1)$ gauge field in the (quasi) de Sitter limit studied in \cite{Kobayashi:2014zza}. Here we summarize the similarities and differences of these two types of the Schwinger effect:

\begin{itemize}
\item{In the $U(1)$ case, the non-zero VEV of the gauge field breaks spatial isotropy. Therefore, both the Bogoliubov coefficients and the induced current are direction dependent. In particular, the induced current is in the direction of the Abelian Electric field and $\kappa_{\bk}\propto \frac{k_z}{k}$. On the other hand, in our setup, the $SU(2)$ VEV is isotropic and so are our current and Bogoliubov coefficients.}
\item{For the $U(1)$ electric field and a scalar field with charge $e$, we have a significant pair production and a sizable induced current in the strong field limit ($\frac{eE}{H^2}\gg1$). For instance, the exponent of the $\beta_{\bk}$ Bogoliubov coefficient in the $U(1)$ case is \cite{Kobayashi:2014zza}
\Beq
i\kappa_{\bk} - \lvert \mu \rvert \approx \left(\frac{k_z}{k}-1\right)\frac{eE}{H^2} - \frac{m^2/H^2-9/4}{2eE/H^2}\,,
\Eeq
which for modes in the $\hat{z}$-direction and small mass can be negligible and generates an order one $\beta_{\bk}$, hence a significant pair production. Moreover, in the large field regime, the current grows like $(\frac{eE}{H^2})^2$ with the electric field value. On the other hand in our case, the pair production is exponentially suppressed in the $\xp\gg1$ limit regardless of the mass. That is due to the fact that in the $SU(2)$ case, the $\beta_{\lambda}$ coefficients in Eq. \eqref{alpha-beta} always have a negative exponent, i.e.,
$$i\kappa_{\lambda} - \lvert \mu \rvert <0,$$
regardless of the parameters. In the limit of $\xp\gg 1$ and $\xp \gg \frac{m}{H}$, by increasing the $\xp$ value, the current decreases like $1/\xp$ and the conductivity like $1/\xp^2$.}
\item{In the small field limit of $\frac{eE}{H^2}\ll \frac{m}{H}$ for the $U(1)$ and $\xp\ll \frac{m}{H}$ for the $SU(2)$ setup, both systems have the similar current and conductivity behaviors. In particular, in this regime, the current grows linearly with the field strength and the conductivity is almost a constant.}
\item{Unlike for the $U(1)$ case, in the $SU(2)$ setup, the induced current has an absolute maximum value $\mathcal{J}^{\rm Max}_{\rm reg} \sim \frac{7}{20} \frac{\ga H^3}{(2\pi)^2}\frac{1}{2m}$ at $\xp\sim 2 \frac{m}{H}$.}
\item{Refs. \cite{Kobayashi:2014zza} and \cite{Hayashinaka:2016qqn} study the scalar and fermion Schwinger effects by a $U(1)$ field and show that there are regions in the parameter space in which the conductivity is negative. These parameters make $\Omega^2_{\bk}=\omega^2_{\bk}+\frac{a''}{a}$ and hence the physical momentum zero. On the other hand, in our $SU(2)$ setup, $\Omega_{\bk}^2$ is positive definite throughout the parameter space and the conductivity is always positive. }
\item{These two setups are different even in the Minkowski limit. The $U(1)$ case leads to the standard Schwinger effect in the flat space. However, the $SU(2)$ gauge field with an isotropic and homogeneous field configuration has zero Schwinger pair production in the Minkowski limit.}
\end{itemize}


\section{Backreaction Constraints on Parameter Space}
\label{sec:Constr}

\begin{figure*}[t] 
   \centering
  \includegraphics[height=4.2in]{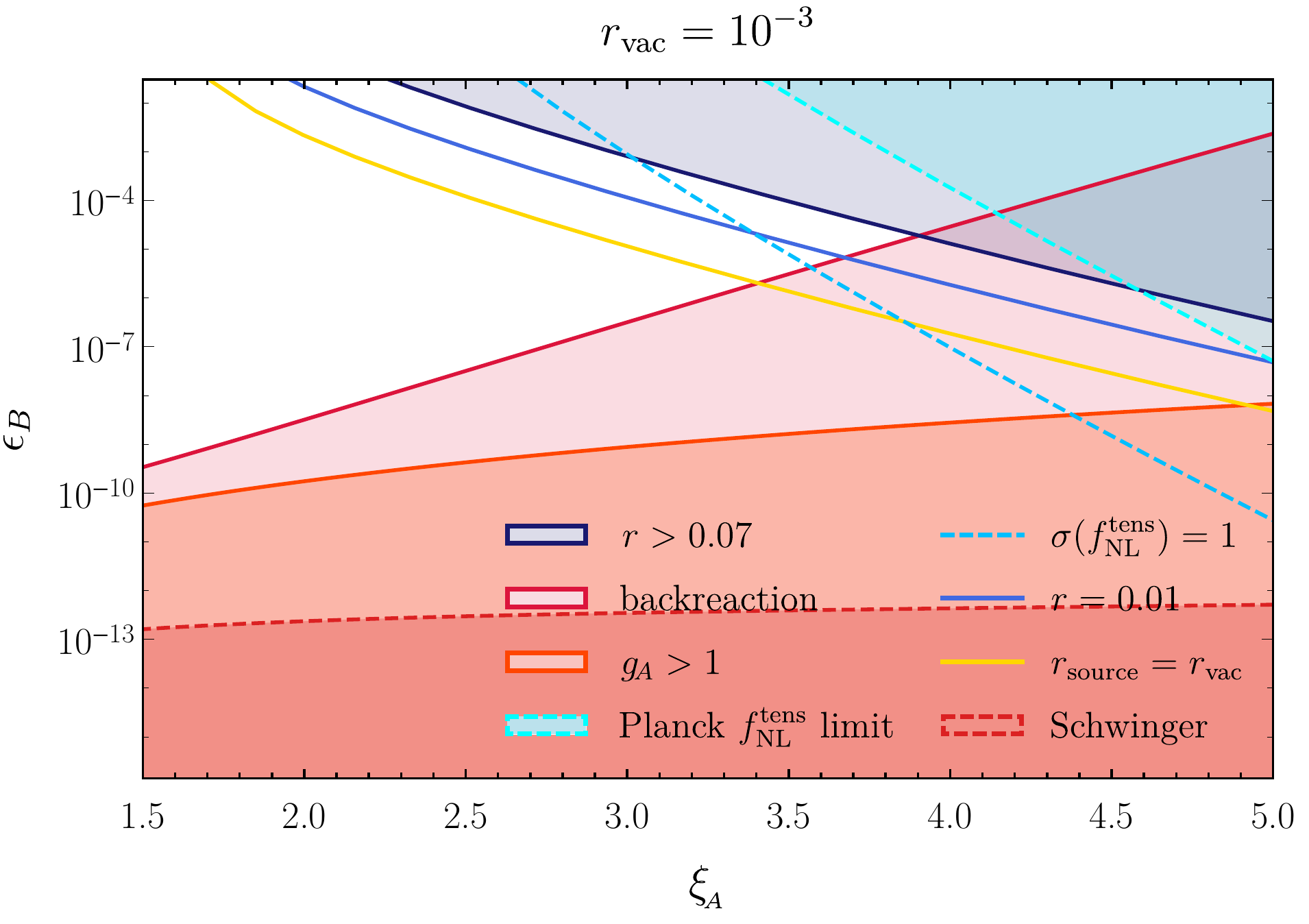} 
   \caption{Various observational and theoretical constraints on the model parameter space, cf. Ref. \cite{Agrawal:2018mrg}. The bound from the Schwinger particle production, depicted by the red dashed line (for $m=0$) and the area underneath, does not lead to additional bounds on the observationally relevant parameter space. The $\xp$ on the horizontal axis is the same as $m_Q$ in Refs. \cite{Dimastrogiovanni:2016fuu,Agrawal:2017awz, Agrawal:2018mrg}.}
   \label{fig:EpsBrvace-3}   
\end{figure*}

Up to now, we studied the scalar Schwinger effect in the presence of the (isotropic) $SU(2)$ scalar gauge field VEV. During slow-roll inflation, these results are independent of the details of the inflationary model that generates the gauge fields. In this section, we study the importance of the backreaction of the induced current on the dynamics of the background $SU(2)$ gauge field, $Q$, for the model given in Eq. \eqref{the-spec}. Specifically, we assume that the background gauge field experiences strong backreaction when the last two terms in Eq. \eqref{eq:QEoM} are at least comparable
\Beq
\label{eq:backr}
\left|\frac{g_{\!A}\lambda_A}{af}\partial_{\tau}\chi Q^2\right|\lesssim \mathcal{J}_{\rm{reg}}\,.
\Eeq
Since $g_{\!A}\lambda_A\partial_{\tau}\chi/(af)\approx2H(\xp+\xp^{-1})$ \cite{Dimastrogiovanni:2016fuu}, Eq. \eqref{eq:backr} is equivalent to
\Beq
\label{eq:SchwEpsB}
\epsilon_B\lesssim\left(\frac{1}{\xp^2}+\frac{1}{\xp^3}\right)^{-1}\frac{\pi^2A_{\rm{s}}r_{\rm{vac}}}{4}\frac{\mathcal{J}_{\rm{reg}}}{g_{\!A}H^3}\,,
\Eeq
where $\epsilon_B\equiv \xp^4H^2/(g_{\!A}^2m_{\rm{pl}}^2)$ and is approximately twice the energy density fraction of the gauge field. Here, we have used the standard single-field slow-roll inflation relation $r_{\rm{vac}}=2H^2/(A_{\rm{s}}\pi^2m_{\rm{pl}}^2)$ to parametrize the energy scale of inflation, i.e., $r_{\rm vac}$ is the vacuum contribution to the tensor-to-scalar ratio in the absence of matter. The measured amplitude of the curvature power spectrum is $A_{\rm{s}}\approx2.2\times10^{-9}$ \cite{Ade:2015xua}. The dashed red line and the shaded area underneath in Fig. \ref{fig:EpsBrvace-3} show the inequality in Eq. \eqref{eq:SchwEpsB}, i.e., regions with strong backreaction due to the induced current. Since $\mathcal{J}_{\rm{reg}}$ decreases monotonically with the mass (see Fig. \ref{fig:JmQ}), $m=0$ gives the lower bound on $\epsilon_B$. We also find that $\epsilon_B$ scales linearly with $r_{\rm{vac}}$ and therefore the Schwinger particle production becomes increasingly less important as the energy scale of inflation is reduced.

The $SU(2)$ backgrounds can source tensor gauge modes, which can also lead to backreaction. Effects on the homogeneous dynamics can arise if \cite{Agrawal:2018mrg}
\Beq
\label{eq:TensorEpsB}
\epsilon_B\lesssim \xp^2\left[\mathcal{B}(\xp)+\frac{\tilde{\mathcal{B}}(\xp)}{\xp+\xp^{-1}}\right]\frac{A_{\rm{s}}r_{\rm{vac}}}{48}\,,
\Eeq
where 
\Beq
\mathcal{B}(\xp)=\int_0^{x_{\rm{max}}}dxx|i^{\beta}W_{\beta,\alpha}(-2ix)|^2\,,\qquad\tilde{\mathcal{B}}(\xp)=\int_0^{x_{\rm{max}}}dxx^2|i^{\beta}W_{\beta,\alpha}(-2ix)|^2\,,
\Eeq
and
\Beq
\alpha=-i\sqrt{2\xp^2+\frac{7}{4}}\,,\quad\beta=-i\sqrt{2\xp+\frac{1}{\xp}}\,,\\
x_{\rm{max}}=2\xp+\frac{1}{\xp}+\sqrt{2\xp^2+2+\frac{1}{\xp^2}}\,.
\Eeq
The solid magenta line and the area underneath in Fig. \ref{fig:EpsBrvace-3} show the inequality given in Eq. \eqref{eq:TensorEpsB}, i.e., regions with strong backreaction due to the tensor gauge field modes production. It fully covers the backreaction region of the Schwinger particle production. Since the strength of the tensor gauge mode backreaction also scales linearly with $r_{\rm{vac}}$, it remains the dominant factor for constraining parameter space for all energy scales of inflation.

We can also infer for what self-coupling constants, $g_{\!_A}$, the Schwinger effect becomes important. Given the parameter relationship
\Beq
\label{eq:EpsBgA1}
\epsilon_B=\frac{\pi^2A_{\rm{s}}r_{\rm{vac}}}{2g_{\!_A}^2}\xp^4\,,
\Eeq
we can draw the line $g_{\!_A}=1$ separating the strongly and weakly coupled regions in $\epsilon_B-\xp$ space. In Fig. \ref{fig:EpsBrvace-3} the orange solid line and the shaded area underneath show the strongly coupled regions, $g_{\!_A}>1$. There we expect non-negligible radiative corrections, i.e., loop contributions, to the gauge field dynamics due to interactions with the scalar field. Such effects could alter the $SU(2)$ dynamics. Since we find that the Schwinger production backreacts on the background dynamics only for $g_{\!_A}\gg1$, we cannot be certain when exactly the scalar excitations affect the $SU(2)$ evolution significantly. However, we can say with absolute certainty that they do not affect the $SU(2)$ dynamics in the weakly coupled regime, $g_{\!_A}<1$. This applies for all energy scales of inflation, since the right hand side in Eq. \eqref{eq:EpsBgA1} again scales linearly with $r_{\rm{vac}}$.

We also conclude that the Schwinger effect can be ignored when the sourced gravitational waves by the $SU(2)$ field have a substantial power spectrum and/or bispectrum. The dark blue solid line in Fig. \ref{fig:EpsBrvace-3} is the current upper bound on the tensor power spectrum \cite{Array:2015xqh}. The light blue, $r_{\rm source}=0.01$, and yellow, $r_{\rm vac}=r_{\rm source}=0.001$, solid lines show different values of the total tensor-to-scalar ratio. In the region left of the yellow line, the sourced gravitational waves have lower power than the vacuum fluctuations in the metric. The light and dark blue dashed lines in Fig. \ref{fig:EpsBrvace-3} show the current and target constraints on the amplitude of tensor non-Gaussianities, $f_{\rm{NL}}^{\rm{tens}}$ \cite{Agrawal:2018mrg}, respectively. The estimated Schwinger bound lies well below any of these observationally interesting regions.


\section{Conclusions}
\label{sec:concl}

In this paper, we have studied the Schwinger effect of a charged scalar doublet by an (isotropic and homogeneous) $SU(2)$ gauge field during inflation. We analytically derived the explicit form of the induced current and Schwinger pair production. We found that the Schwinger effect by the $SU(2)$ gauge field is very different from that by a (homogeneous) Abelian gauge field with a preferred spatial direction. We showed that these two cases are very different even in the Minkowski limit. In particular, the $SU(2)$ gauge field VEV adds an extra mass term to the scalar field in such a way that this setup has negligible particle production even in the strong field limit. That is unlike the standard Abelian Schwinger effect in which a complex scalar field with a small mass experiences a significant pair production in the strong field limit \cite{Kobayashi:2014zza}. The $\xp\approx2m/H$ point in parameter space divides the behavior of the induced scalar current into two different regimes. For $\xp<2m/H$, the current increases linearly with $\xp$, while for $\xp>2m/H$ it decreases like $\xp^{-1}$. Its maximum value at $\xp\approx2m/H$ is $\mathcal{J}^{\rm max}\approx5\times10^{-3}g_{\!_A}H^3/m$.

Finally, we used our result to study the possible backreaction of the induced current on the non-Abelian background dynamics and to further constrain an axion-$SU(2)$ inflation model. We found that the backreaction by the induced current is important only for cases when backreaction due to tensor gauge mode amplification is also significant and the self-coupling of the $SU(2)$ fields is strong, $g_{\!_A}\gg1$. Therefore, the Schwinger production of charged scalar fields by an $SU(2)$ field does not yield additional constraints on the observationally relevant parameter space. 

We expect that our conclusions can be extended to charged spin-$1/2$ fields, since the effect will be even weaker due to suppression in particle production by the Pauli exclusion principle. A rigorous proof is left for future work. The isotropy of the $SU(2)$ gauge field played an important role in suppressing the particle production and the decrease of the induced current in the strong field regime. Therefore, an important question is a possible effect of statistical anisotropy on the Schwinger process by an $SU(2)$ gauge field during inflation which we also postpone for future work.


\acknowledgments

We are grateful to Aniket Agrawal and Takeshi Kobayashi for helpful discussions and to Giovanni Cabass for a careful proof-reading. EK thanks Raphael Flauger for the question he asked during the `Fundamental Cosmology' meeting held at Teruel from September 11th to 13th, 2017, which led to this project. He also thanks Emanuela Dimastrogiovanni, Matteo Fasiello, and Tomohiro Fujita for useful discussions.


\appendix

\section{Computation of the total current}
\label{Current-appendix}

In this appendix, we present the computation of the current, $\mathcal{J}$. The current due to a constant electric field (with a preferred spatial direction) has been worked out in \cite{Kobayashi:2014zza}. Comparing to the $U(1)$ case, our current integrand enjoys spatial isotropy and the details of $\kappa$ and $\mu$ coefficients are different and direction independent. The expectation value of the current given in \eqref{curly-J} can be written as
\be\label{cJ-to-cG}
\langle 0_{\rm in}\lvert J^a_{i} \rvert 0_{\rm in}\rangle =  \delta^a_i a\mJ = \frac{ \ga aH^3}{(2\pi)^2}  \delta^a_i \sum_{\lambda=\pm1} \mathcal{G}_{\lambda},
\ee 
where $\mathcal{G}_{\lambda}$ is the following integral 
\be\label{mJ-appendix}
\mathcal{G}_{\lambda}(\Lambda) =  \lim_{\Lambda \rightarrow\infty} e^{i  \kl \pi} \int^{\Lambda}_0 \x'd\x' \big( - \frac13 \lambda \x' + \frac12 \xp \big) \big\lvert W_{\kappa_{\lambda},\mu}(-2i\x')\big\rvert^2.
\ee
Here $\x$ is a rescaled physical momentum and $\Lambda$ is the UV cutoff on $\x$
 $$\x \equiv \frac{k}{aH} \an  \Lambda\equiv\frac{k_{\rm{UV}}}{aH}\,.$$ 
 We shall send $\Lambda$ to infinity in the end. In order to perform the integral in \eqref{mJ-appendix}, we use the Mellin-Barnes integral representation of the Whittaker functions \cite{Nist}
\be\label{Mellin-Barnes}
W_{\kappa,\mu}(z) = \frac{e^{-\frac{z}{2}}}{2i\pi} \int^{i\infty}_{-i\infty} \frac{\Gamma(\frac12+\mu+s)\Gamma(\frac12-\mu+s)\Gamma(-\kappa-s)}{\Gamma(\frac12+\mu-\kappa)\Gamma(\frac12-\mu-\kappa)} z^{-s}ds  \where \lvert {\rm{arg}}(z)\rvert <\frac32\pi, \nonumber\\
\ee 
which holds when
\be
\frac12\pm \mu -\kappa \neq 0, -1, -2 ,\dots,  \nonumber
\ee
and the contour of the integration separates the poles of $\Gamma(\frac12+\mu+s)
\Gamma(\frac12-\mu+s)$ at \footnote{The $\Gamma(z)$ function has simple poles for non-positive values of $z$ at
\be\label{Gamma-r}
z=-n \quad \forall n\in \mathbb{N} \where {\rm{Res}}(\Gamma,-n)= \frac{(-1)^n}{n!}.
\ee}
\be\label{snpm}
s_{n,\pm}=-\frac12-n\pm \mu \quad \forall n\in \mathbb{N},
\ee
and the poles of $\Gamma(-\kappa-s)$ at
\be\label{sn0}
s_{n,0}=-\kappa+n \quad \forall n\in \mathbb{N}.
\ee 
In our setup, $\kappa$ is pure imaginary and $\mu$ is either real or pure imaginary (i.e., $\kappa^{*}=-\kappa$ and $\mu^{*}$ is either $\mu$ or $-\mu$). Moreover, from \eqref{kappa-mu} and recalling the fact that the consistency of our model requires $\xp>\sqrt{2}$, we have $0\leq \rm{Re}(\mu)< \frac{\sqrt{3}}{2}$.

\begin{figure*}[t] 
   \centering
     \includegraphics[width=2.5in]{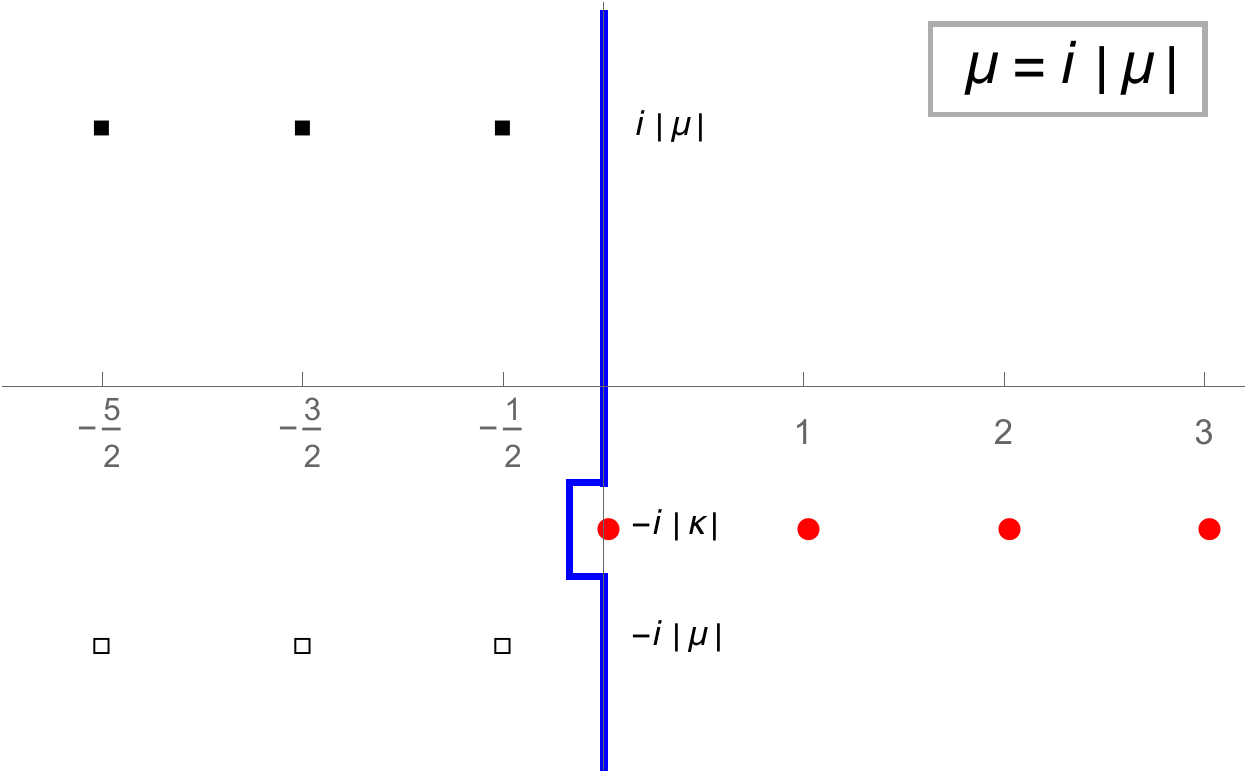} 
  \hspace{0.05in}
{\includegraphics[width=2.5in]{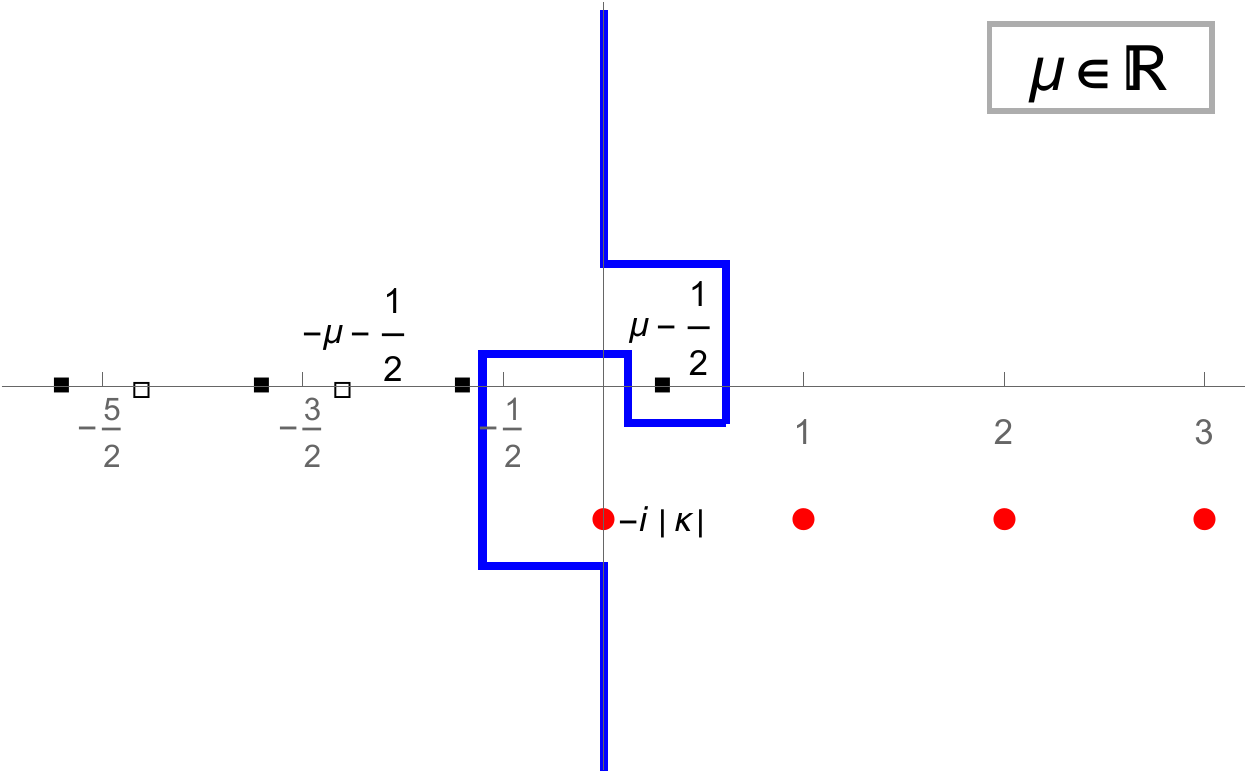}}
   \caption{Positions of poles $s_{n,\pm}$ and $s_{n,0}$ and the possible contouring for the validity of the Mellin-Barnes integral representation in \eqref{Mellin-Barnes} are shown for the $\lambda=-$ polarization state with $\kappa_{-}=i \lvert \kappa \rvert$. In the left panel, we show the locations of poles for purely imaginary values of $\mu$, while in the right panel for purely real values of $\mu$. The (red) circles and the black and white squares show poles of type $s_{n,0}$, $s_{n,+}$ and $s_{n,-}$ respectively. For the $\lambda=+$ polarization state, we have the same poles, but for $\kappa_{+}=-i \lvert \kappa \rvert$. The Mellin-Barnes integral is valid for any closed contour as far as it separates the circle poles from the square poles and it runs from $-i\infty$ to $+i\infty$. We show one possible contouring on the imaginary axis. The contour can be closed in either the left or right half-plane. }
   \label{fig:poles}   
\end{figure*}

Using \eqref{Mellin-Barnes} and  doing the $\x'$ integral , we arrive at
\be\label{mG-1}
&& \mathcal{G}_{\lambda}(\Lambda) =  \lim_{\Lambda \rightarrow\infty}  \frac{1}{24(2\pi)^2} e^{i \kappa_{\lambda}\pi} \bigg[\Gamma(\frac12+\mu-\kappa)\Gamma(\frac12-\mu-\kappa)\Gamma(\frac12+\mu^*+\kappa)\Gamma(\frac12-\mu^*+\kappa)
\bigg]^{-1}\nonumber\\
&&  \int^{i\infty}_{-i\infty} ds \int^{i\infty}_{-i\infty} ds' e^{i(s-s')\frac{\pi}{2}}(2\Lambda)^{2-s-s'}\bigg( -\frac{2\lambda\Lambda}{3-s-s'} + \frac{3\xp}{2-s-s'}\bigg) \Gamma(\frac12+\mu+s)\Gamma(\frac12-\mu+s) \nonumber\\
&& \Gamma(-\kappa-s) \Gamma(\frac12+\mu^{*}+s')\Gamma(\frac12-\mu^*+s')\Gamma(\kappa-s').
\ee
Thus, the complex integrand of $s'$-integral has singularities at $s'=-\frac12 \pm \mu  -n$ and $s'=\kappa+n$ as well as at $s'=3-s$ and $s'=2-s$. Moreover, since the integrand is proportional to $\Lambda^{2-s-s'}$ in the limit that $\Lambda \rightarrow\infty$, it vanishes for ${\rm{Re}}(s')>3-{\rm{Re}}(s)$. Upon choosing the contour of $s$ such that ${\rm{Re}}(s)>-1$ and closing the $s'$-contour in the right-half plane without passing through the poles,  we are left with the following six poles
\be
s'_1=\kappa, ~ s'_2=\kappa+1, ~s'_3=\kappa+2, ~s'_3=\kappa+3, ~s'_5=2-s \an s'_6=3-s.
\ee
Notice that an integral of \eqref{mG-1} over a finite path along the real axis vanishes at $\lim \rm{Im}(s') \rightarrow \pm \infty$ (due to the Gamma functions) and therefore the added integration contour does not change the result.  
Using the Cauchy integral theorem, we can do the $s'$-integral
\be
&& G_{\lambda}(s,\Lambda) \equiv \nonumber\\
&& \lim_{\Lambda\rightarrow \infty} \int^{i\infty}_{-i\infty} ds' e^{i(s-s')\frac{\pi}{2}}(2\Lambda)^{2-s-s'}\bigg( \frac{2\lambda\Lambda}{s'+s-3} - \frac{3\xp}{s'+s-2}\bigg) 
\Gamma(\frac12+\mu^{*}+s')\Gamma(\frac12-\mu^*+s')\Gamma(\kappa-s')\,.  \nonumber
\ee
Using \eqref{Gamma-r}, we obtain 
\be\label{G}
&&  G_{\lambda}(s,\Lambda)= \nonumber \\
&& \lim_{\Lambda\rightarrow \infty}(2i\pi) e^{-i\kappa \pi}\bigg[ e^{i(s+\kappa)\frac{\pi}{2}}(2\Lambda)^{2-s-\kappa}\bigg(\frac{2\lambda\Lambda}{s+\kappa-3}- \frac{3\xp}{s+\kappa-2}\bigg) 
\Gamma(\frac12+\mu^{*}+\kappa)\Gamma(\frac12-\mu^*+\kappa)  \nonumber \\
&& -~~~ e^{i(s+\kappa-1)\frac{\pi}{2}}(2\Lambda)^{1-s-\kappa}\bigg( \frac{2\lambda\Lambda}{s+\kappa-2} - \frac{3\xp}{s+\kappa-1}\bigg) 
\Gamma(\frac32+\mu^{*}+\kappa)\Gamma(\frac32-\mu^*+\kappa)   \nonumber \\
&& +~~~\frac{1}{2!} e^{i(s+\kappa-2)\frac{\pi}{2}}(2\Lambda)^{-s-\kappa}\bigg( \frac{2\lambda\Lambda}{s+\kappa-1} - \frac{3\xp}{s+\kappa}\bigg) 
\Gamma(\frac52+\mu^{*}+\kappa)\Gamma(\frac52-\mu^*+\kappa) \nonumber \\
&& -~~~\frac{1}{3!} e^{i(s+\kappa-3)\frac{\pi}{2}}(2\Lambda)^{-s-\kappa} \frac{\lambda}{s+\kappa} 
\Gamma(\frac72+\mu^{*}+\kappa)\Gamma(\frac72-\mu^*+\kappa)  
 \nonumber\\
&& - ~~~ 3\xp e^{i(s+\kappa-1)\pi} \Gamma(\frac52+\mu^{*}-s)\Gamma(\frac52-\mu^*-s)\Gamma(\kappa-2+s) \nonumber\\
&& + ~~~ \lambda e^{i(s+\kappa-\frac32)\pi} \Gamma(\frac72+\mu^{*}-s)\Gamma(\frac72-\mu^*-s)\Gamma(\kappa-3+s)\bigg].
\ee
This integral is divergent and includes terms proportional to $\Lambda^3$, $\Lambda^2$ and $\Lambda$. However, the last two terms are finite and independent of $\Lambda$.  Now we turn to compute the last complex integral with respect to $s$ which is
\be\label{tG-tot}
&& \mathcal{G}_{\lambda}(\Lambda)  =  \frac{e^{i\kappa\pi}}{24(2\pi)^2} \bigg[\Gamma(\frac12+\mu-\kappa)\Gamma(\frac12-\mu-\kappa)\Gamma(\frac12+\mu^*+\kappa)\Gamma(\frac12-\mu^*+\kappa)
\bigg]^{-1}
\nonumber\\
&& \times \lim_{\Lambda\rightarrow \infty}\int^{i\infty}_{-i\infty} ~ds ~ \Gamma(\frac12+\mu+s)\Gamma(\frac12-\mu+s)  \Gamma(-\kappa-s)  G_{\lambda}(s,\Lambda).
\ee
In doing the integral of the four $\Lambda$-dependent terms, we close the contour in the right-half plane. Here, we have the following poles in the $\Lambda\rightarrow\infty$ limit
$$s_1=-\kappa, ~s_2=1-\kappa, ~ s_3=2-\kappa \an s_4=3-\kappa,$$
and, unlike the first complex integral, some of them are not simple, but of order 2. \footnote{Here we use the following formula in the complex analysis that if $f(z)$
has a pole of order $k$ at $z = z_0$ then the residues are given as
\be
{\rm{Res}}(f,z_0)= \frac{1}{(k-1)!} \frac{d^{k-1}}{dz^{k-1}}\bigg( (z-z_0)^k f(z)\bigg)\bigg\rvert_{z=z_0}.
\ee}

In order to compute $\mathcal{G}_{\lambda}(\Lambda)$ in \eqref{tG-tot}, first we compute the second complex integral for each line of $G_{\lambda}(s,\Lambda)$ in \eqref{G} individually. Specifically, we label the integral corresponding to the $I$-th line in \eqref{G} as $\mathcal{G}_{\lambda,I}(\Lambda)$, where $I=1,2,...6$. in the end, we sum up all the $\mathcal{G}_{\lambda,I}(\Lambda)$s and determine $\mathcal{G}_{\lambda}(\Lambda)$.
Doing the second complex integral in \eqref{tG-tot}, the first line of $ G_{\lambda}(s,\Lambda)$ in \eqref{G} leads to the following $\mathcal{G}_{\lambda,1}(\Lambda)$:
\be\label{tG1}
&& \mathcal{G}_{\lambda,1}(\Lambda)=\nonumber\\
&& \frac{1}{24}\lim_{\Lambda\rightarrow \infty} \bigg( \Gamma(\frac12+\mu-\kappa)\Gamma(\frac12-\mu-\kappa) \bigg)^{-1}
 \bigg[ (2\Lambda)^{2}\big( -\frac{2\lambda\Lambda}{3} + \frac{3\xp}{2}\big) 
\Gamma(\frac12+\mu-\kappa)\Gamma(\frac12-\mu-\kappa)  \nonumber \\
&& +(-1)e^{i\frac{\pi}{2}}(2\Lambda)\big( -\lambda\Lambda + 3\xp\big)\Gamma(\frac32+\mu-\kappa)\Gamma(\frac32-\mu-\kappa)  -\frac{(-1)^2}{2!}e^{i\pi} \lambda (2\Lambda) 
\Gamma(\frac52+\mu-\kappa)  \nonumber \\
&& \Gamma(\frac52-\mu-\kappa) 
 + 3\xp \frac{d}{ds}\bigg((s+\kappa-2)e^{i(s+\kappa)\frac{\pi}{2}}(2\Lambda)^{2-s-\kappa}\Gamma(\frac12+\mu+s)\Gamma(\frac12-\mu+s)  \Gamma(-\kappa-s) \bigg)_{s=2-\kappa}
  \nonumber \\
&& - \lambda  \frac{d}{ds}\bigg((s+\kappa-3)e^{i(s+\kappa)\frac{\pi}{2}}(2\Lambda)^{3-s-\kappa}\Gamma(\frac12+\mu+s)\Gamma(\frac12-\mu+s)  \Gamma(-\kappa-s) \bigg)_{s=3-\kappa}
\bigg].  \nonumber
\ee
This integral includes a derivative of the $\Gamma$-function which can be written as a digamma function 
\be
\psi(z) =\frac{\p_{z} \Gamma(z)}{\Gamma(z)}.
\ee
In particular, the second order roots in \eqref{tG-tot} leads to the derivative terms of the form \footnote{Here we used the fact that the harmonic series is $\sum_{q=1}^n \frac{1}{q}= -\psi(1) +\psi(n+1)$, and the following relations 
$$\lim_{z \rightarrow0} z \Gamma(-z-n) =- \frac{(-1)^n}{n!} \an \lim_{z \rightarrow0} \frac{d}{dz}(z \Gamma(-z-n))=  \frac{(-1)^n}{n!} \psi(n+1) \quad \forall n\in \mathbb{N}.$$
}
\be\label{del}
&& -\frac{d}{ds}\bigg[(s+\kappa-n)e^{i(s+\kappa)\frac{\pi}{2}}(2\Lambda)^{n-s-\kappa}\Gamma(\frac12+\mu+s)\Gamma(\frac12-\mu+s)  \Gamma(-\kappa-s) \bigg]\bigg\rvert_{s+\kappa=n}   \nonumber\\
&& = \frac{(-1)^{n}}{n!} e^{i\frac{n\pi}{2}}
\Gamma(\frac{1+2n}{2}+\mu-\kappa)\Gamma(\frac{1+2n}{2}-\mu-\kappa) 
\bigg[\frac{i\pi}{2}-\ln(2\Lambda) -\psi(n+1) \nonumber\\
&& +  \psi(\frac{1+2n}{2}+\mu-\kappa)   +\psi(\frac{1+2n}{2}-\mu-\kappa) \bigg],
\ee
when $n$ is an integer. 
Using \eqref{del}, we obtain $\mathcal{G}_{\lambda,1}$ as
\be
&& \mathcal{G}_{\lambda,1}(\Lambda)= \frac{1}{24}\lim_{\Lambda\rightarrow \infty} 
 \bigg\{ (2\Lambda)^{2}\bigg[ -\frac{2\lambda\Lambda}{3} + \frac{3\xp}{2}\bigg] 
 -i(2\Lambda)\bigg[ -\lambda\Lambda + 3\xp\bigg] \bigg[(\frac12-\kappa)^2-\mu^2\bigg]  
  \nonumber \\
 && +\frac{\lambda}{2} (2\Lambda)   \bigg[(\frac12-\kappa)^2-\mu^2\bigg]  \bigg[(\frac32-\kappa)^2-\mu^2\bigg]
  \nonumber \\
&& 
+ \frac{3}{2} \xp  \bigg[(\frac12-\kappa)^2-\mu^2\bigg]  \bigg[(\frac32-\kappa)^2-\mu^2\bigg]
\bigg[\frac{i\pi}{2}-\ln(2\Lambda) -\psi(3) +  \psi(\frac{5}{2}+\mu-\kappa)+\psi(\frac{5}{2}-\mu-\kappa) \bigg]
  \nonumber \\
&& + \frac{i\lambda}{6} 
 \bigg[(\frac12-\kappa)^2-\mu^2\bigg]  \bigg[(\frac32-\kappa)^2-\mu^2\bigg] \bigg[(\frac52-\kappa)^2-\mu^2\bigg] 
\bigg[\frac{i\pi}{2}-\ln(2\Lambda) -\psi(4) \nonumber\\
&&+  \psi(\frac{7}{2}+\mu-\kappa)+\psi(\frac{7}{2}-\mu-\kappa) \bigg]
\bigg\}.  \nonumber
\ee

Next, doing the $s$-integral and using \eqref{del}, we find $\mathcal{G}_{\lambda,2}$ as 
\be
&& \mathcal{G}_{\lambda,2}(\Lambda)= \lim_{\Lambda\rightarrow \infty}  \frac{1}{24}   \bigg[(\frac12+\kappa)^2-\mu^2\bigg] \bigg\{ i(2\Lambda) \big( -\lambda\Lambda + 3\xp\big)  - \lambda(2\Lambda) \bigg[(\frac12-\kappa)^2-\mu^2\bigg] 
  \nonumber \\
&& - 3 \xp \bigg[(\frac12-\kappa)^2-\mu^2\bigg] 
\bigg[\frac{i\pi}{2}-\ln(2\Lambda) -\psi(2) +  \psi(\frac{3}{2}+\mu-\kappa)+\psi(\frac{3}{2}-\mu-\kappa) \bigg]
  \nonumber \\
&& - \frac{i\lambda}{2}  \bigg[(\frac12-\kappa)^2-\mu^2\bigg] \bigg[(\frac32-\kappa)^2-\mu^2\bigg] 
\bigg[\frac{i\pi}{2}-\ln(2\Lambda) -\psi(3) +  \psi(\frac{5}{2}+\mu-\kappa)+\psi(\frac{5}{2}-\mu-\kappa) \bigg]\bigg\},  \nonumber
\ee
while $\mathcal{G}_{\lambda,3}$ is
\be
&& \mathcal{G}_{\lambda,3}(\Lambda)=  \lim_{\Lambda\rightarrow \infty} \frac{1}{24} \bigg[(\frac12+\kappa)^2-\mu^2\bigg] \bigg[(\frac32+\kappa)^2-\mu^2\bigg]  \bigg\{ \lambda  \Lambda  \nonumber \\
&&  + \frac32 \xp \bigg[\frac{i\pi}{2}-\ln(2\Lambda) -\psi(1) +  \psi(\frac{1}{2}+\mu-\kappa)+\psi(\frac{1}{2}-\mu-\kappa) \bigg] \nonumber\\
&&  + \frac{i}{2} \lambda  \bigg[(\frac12-\kappa)^2-\mu^2\bigg] 
\bigg[\frac{i\pi}{2}-\ln(2\Lambda) -\psi(2) +  \psi(\frac{3}{2}+\mu-\kappa)+\psi(\frac{3}{2}-\mu-\kappa) \bigg]
\bigg\}, \nonumber
\ee
and $\mathcal{G}_{\lambda,4}$ is
\be
&& \mathcal{G}_{\lambda,4}(\Lambda)=  \lim_{\Lambda\rightarrow \infty} - \frac{i \lambda}{3!}\frac{1}{24}   \bigg[(\frac12+\kappa)^2-\mu^2\bigg]\bigg[(\frac32+\kappa)^2-\mu^2\bigg]\bigg[(\frac52+\kappa)^2-\mu^2\bigg]  \nonumber\\
&& \times   \bigg[\frac{i\pi}{2}-\ln(2\Lambda) -\psi(1) +  \psi(\frac{1}{2}+\mu-\kappa)+\psi(\frac{1}{2}-\mu-\kappa) \bigg].
\ee

Now we turn to compute the last two ($\Lambda$-independent) terms in \eqref{G} and find $\mathcal{G}_{\lambda,5}$ and $\mathcal{G}_{\lambda,6}$. Using the following relation for $\Gamma$ functions 
\be
\Gamma(z)\Gamma(-z) = -\frac{\pi}{z \sin(\pi z)}    \quad (z\notin \mathbb{Z}),
\ee 
we find that $\mathcal{G}_{\lambda,5}+\mathcal{G}_{\lambda,6}$ can be written as 
\be
\mathcal{G}_{\lambda,5}+\mathcal{G}_{\lambda,6} = \frac{1}{24}  \bigg[\Gamma(\frac12+\mu-\kappa)\Gamma(\frac12-\mu-\kappa)\Gamma(\frac12+\mu^*+\kappa)\Gamma(\frac12-\mu^*+\kappa)
\bigg]^{-1}  \mathcal{I}_{\lambda},
\ee
where 
\be
\mathcal{I}_{\lambda} &=&  \frac{ \pi }{(2i\pi)} \int_{-i\infty}^{+i\infty} ds ~  \frac{e^{i(s+\kappa)\pi}}{\sin(\pi(s+\kappa))}  \Gamma(\frac12+\mu^{*}-s)\Gamma(\frac12-\mu^*-s)\Gamma(\frac12+\mu+s)\Gamma(\frac12-\mu+s)\nonumber\\
&&  \frac{\big((\frac32-s)^2-\mu^2\big)\big((\frac12-s)^2-\mu^2\big)}{(\kappa+s)(\kappa+s-1)(\kappa+s-2)}  \bigg[3\xp + i\lambda \frac{\big((\frac52-s)^2-\mu^2\big)}{(\kappa+s-3)} \bigg].
\ee 
The integrand has an infinite number of poles. Therefore, it is more convenient to express it in the following form 
\be\label{G5-G6}
\mathcal{I}_{\lambda}  & = &  \frac{\pi}{(2i\pi)}  \int_{-i\infty}^{+i\infty} ds ~  \Gamma(\frac12+\mu^{*}-s)\Gamma(\frac12-\mu^*-s)\Gamma(\frac12+\mu+s)\Gamma(\frac12-\mu+s) \frac{ e^{i(s+\kappa)\pi}}{\sin(\pi(\kappa+s))}  \nonumber\\
&& \times  \bigg[ w(s)-w(s-1) + \frac{\mathcal{A}(\kappa,\mu)}{(\kappa+s)} \bigg],
\ee
where 
\be
\mathcal{A} = && 3\xp(\frac12+6\kappa^2-2\mu^2)-i\lambda(7+20\kappa^2-12\mu^2)\kappa  = -2(1+\xp^2)\xp, \\ 
 w = && \frac{3f(s+1)+f(s+3)-3f(s+2)+2g(s+1)-g(s+2)}{\kappa+s} \nonumber\\
&&+  \frac{f(s+2)-3 f(s+1)-g(s+1)}{\kappa+s-1} + \frac{f(s+1)}{\kappa+s-2}\nonumber\\
&& +9\xp s(s+1-2\kappa)  +i\lambda(7+20\kappa^2-12\mu^2)s \nonumber\\
&&+ \frac{10i}{3}\lambda s(s+1)(1+2s-3\kappa),
\ee
in which $g(s)$ and $f(s)$ are
\be
g(s)&=&\frac{3\xp}{2} \bigg[(\frac32-s)^2-\mu^2\bigg]\bigg[(\frac12-s)^2-\mu^2\bigg]\,,\\
f(s)&=& -\frac{i\lambda }{6} \bigg[(\frac52-s)^2-\mu^2\bigg] \bigg[(\frac32-s)^2-\mu^2\bigg]\bigg[(\frac12-s)^2-\mu^2\bigg]\,.
\ee

The first two terms in integral \eqref{G5-G6} can be written as \footnote{Here we used the fact that $\mu$ is either pure imaginary or real (i.e. $\mu^{*}=\pm \mu$) and hence the four Gamma functions in \eqref{I1} are equal to $\frac{\pi^2}{\cos(\pi(\mu-s))\cos(\pi(\mu+s))}$ which is invariant under $s \rightarrow s-1$.}
\be\label{I1-0}
&& \mathcal{I}_{\lambda,1}=  \frac{\pi}{(2i\pi)}\bigg( \int_{-i\infty}^{+i\infty} ds - \int_{-i\infty-1}^{+i\infty-1} ds \bigg) ~\Gamma(\frac12+\mu^{*}-s)\Gamma(\frac12-\mu^*-s)\Gamma(\frac12+\mu+s)\Gamma(\frac12-\mu+s) \nonumber\\
&& \frac{ e^{i(s+\kappa)\pi}}{\sin(\pi(\kappa+s))} w(s).
\ee
The integrand has poles at $s=s_{n,\pm}$ and $s_{n,0}$ (see \eqref{snpm} and \eqref{sn0}) as well as at
$$\tilde{s}_{n,\pm}= \frac12+n \pm \mu \an \tilde{s}_{n,0}= -\kappa-n,$$
in which poles $s_{n,0}$ with $n=0,1,2$ are 2nd rank while the rest are simple poles. Therefore, it is more convenient to close the contour path of $s$ on the left half-plane. Thus, only the simple poles below contribute to the $\mathcal{I}_{\lambda,1}$ integral
$$s_1=-1-\kappa , \quad  s_{0,\pm}=-\frac12 \pm \mu .$$
Note that if $\mu$ is real ($0<\rm{Re}(\mu)<\frac{\sqrt{3}}{2}$), the pole $\tilde{s}_{0,-}=\frac12-\mu$ might be on the left half-plane. However, as it has been shown in figure \ref{fig:poles}, it is always possible to choose the contour path so that $\tilde{s}_{0,-}$ be out of the contour. \footnote{For $\mu=\frac12$ we have $\tilde{s}_{0,-}=s_{0,+}$ and the above contouring is not possible. However, in that case, we can assume $\mu \neq\frac12$ and take the limit $\mu\rightarrow \frac12$.} We obtain

\be\label{I1}
 \mathcal{I}_{\lambda,1} &=&   \Gamma(\frac12+\mu^{*}+\kappa)\Gamma(\frac12-\mu^*+\kappa)\Gamma(\frac12+\mu-\kappa)\Gamma(\frac12-\mu-\kappa) \bigg\{  w(-1-\kappa) \nonumber\\
& + & \frac{i}{2}  \bigg[ \frac{ e^{2i\pi\kappa} + e^{2i\pi\mu}}{\sin(2\pi\mu)} w(-\frac12+\mu) - \frac{e^{2i\pi\kappa} + e^{-2i\pi\mu}}{\sin(2\pi\mu)} w(-\frac12-\mu) \bigg] \bigg\}\,.
\ee
The last term in \eqref{G5-G6} can be written as
\be\label{I2}
\mathcal{I}_{\lambda,2}  & = &  \frac{\mathcal{A}(\kappa,\mu)}{(2i\pi)} \pi \int_{-i\infty}^{+i\infty} ds ~  \Gamma(\frac12+\mu^{*}-s)\Gamma(\frac12-\mu^*-s)\Gamma(\frac12+\mu+s)\Gamma(\frac12-\mu+s) \nonumber \\
&& \times \frac{ e^{i(s+\kappa)\pi}}{(\kappa+s)\sin(\pi(\kappa+s))}\,.
\ee
Closing the contour of the above integral on the left half-plane with an infinite radius semicircle, we have the poles
$$s_{n}=-\kappa-n-1 \an s^{\pm}_{n}=-\frac12 \pm \mu -n, \quad (n=0,1,2,...)$$
 which are an infinite number of simple poles. Technically, in taking the infinite radius limit, we use $(\kappa+s)^{d}~ (d>1)$ in the denominator of \eqref{I2}  and then determine the $d\rightarrow1$ limit solution. Thus the value of the integral over the added infinite radius semicircle vanishes. Doing the above integral and summing up the poles of type $s_{n}$, we have 
\be
\label{I2-}
\mathcal{I}_{\lambda,2}\big\rvert_{s_n-poles} = -  \mathcal{A}(\kappa,\mu) \big( \lim_{d\rightarrow1} \sum_{n=1}^{\infty} \frac{1}{n^{d}}\big) \Gamma(\frac12+\mu^{*}+\kappa)\Gamma(\frac12-\mu^*+\kappa)\Gamma(\frac12+\mu-\kappa)\Gamma(\frac12-\mu-\kappa)\,. \nonumber
\ee
Doing the harmonic series \footnote{Here we used the series representation of a polygamma function
$\psi^{d-1}(z) = (-1)^{d} (d-1)! \sum^{\infty}_{j=0} \frac{1}{(z+j)^{d}},$ which holds for $d > 1$ and any complex $z$ not equal to a negative integer. }, we find 
\be
\label{I2-t1}
\mathcal{I}_{\lambda,2}\big\rvert_{s_n-poles} =  \mathcal{A}(\kappa,\mu)~\psi(1)~ \Gamma(\frac12+\mu^{*}+\kappa)\Gamma(\frac12-\mu^*+\kappa)\Gamma(\frac12+\mu-\kappa)\Gamma(\frac12-\mu-\kappa). \nonumber
\ee
For the poles of type $s^{\pm}_{n}$, we have 
\be
\label{I2-t2}
&& \mathcal{I}_{\lambda,2}\big\rvert_{s^{\pm}_n-poles} =     \frac{i \mathcal{A}(\kappa,\mu)}{2\sin(2\mu\pi)}  \Gamma(\frac12+\mu^{*}+\kappa)\Gamma(\frac12-\mu^*+\kappa)\Gamma(\frac12+\mu-\kappa)\Gamma(\frac12-\mu-\kappa) \nonumber\\
&& \sum^{\infty}_{n=0}  \bigg( \frac{e^{2i\kappa\pi}+e^{2i\mu\pi}}{\kappa+\mu-\frac12-n} - \frac{e^{2i\kappa\pi}+e^{-2i\mu\pi}}{\kappa-\mu-\frac12-n} \bigg),
\ee
which after doing the summation, the term in the second line gives 
\be
\label{I2-t2}
&& \mathcal{I}_{\lambda,2}\big\rvert_{s^{\pm}_n-poles} =    \mathcal{A}(\kappa,\mu)   \Gamma(\frac12+\mu^{*}+\kappa)\Gamma(\frac12-\mu^*+\kappa)\Gamma(\frac12+\mu-\kappa)\Gamma(\frac12-\mu-\kappa) \nonumber\\
&& \frac{i}{2\sin(2\mu\pi)}  \bigg[ (e^{2i\kappa\pi}+e^{2i\mu\pi}) \psi(\frac12-\kappa-\mu) - (e^{2i\kappa\pi}+e^{-2i\mu\pi}) \psi(\frac12-\kappa+\mu)\bigg].
\ee
Therefore, we have
\be\label{I1-}
&& \mathcal{G}_{\lambda,5} +  \mathcal{G}_{\lambda,6}  = \frac{1}{24}   \bigg\{   w(-1-\kappa) +  \frac{i}{2}  \bigg[ \frac{ e^{2i\pi\kappa} + e^{2i\pi\mu}}{\sin(2\pi\mu)} w(-\frac12+\mu) - \frac{e^{2i\pi\kappa} + e^{-2i\pi\mu}}{\sin(2\pi\mu)} w(-\frac12-\mu) \bigg]   \nonumber\\
&& + \mathcal{A}(\kappa,\mu) \bigg[ \psi(1) + \frac{i}{2}  \frac{(e^{2i\kappa\pi}+e^{2i\mu\pi})}{\sin(2\mu\pi)} \psi(\frac12-\kappa-\mu) - \frac{i}{2}  \frac{(e^{2i\kappa\pi}+e^{-2i\mu\pi})}{\sin(2\mu\pi)} \psi(\frac12-\kappa+\mu) \bigg] \bigg\}, \nonumber \\
\ee
in which 
\be
\frac12\bigg[w(-\frac12+\mu) + w(-\frac12-\mu)\bigg] &=& (-1+4\mu^2)\xp + \frac{i\lambda}{2}(-7+12\mu^2-4\xp^2),\\
\frac12\bigg[w(-\frac12+\mu) - w(-\frac12-\mu)\bigg] &=& -\frac{4i\lambda\mu}{3}(-4+4\mu^2-3\xp^2).
\ee
Moreover, summing up $\mathcal{G}_{\lambda,1}$ to $\mathcal{G}_{\lambda,4}$, we find 
\be
\sum_{j=1}^{4} \mathcal{G}_{\lambda,j} &=& - \lim_{\Lambda\rightarrow \infty} \bigg\{ \frac{\lambda}{9} \Lambda^{3} - \frac16 \xp  \Lambda^{2}  +  \frac{\lambda}{3} \Lambda \bigg( 1 - \frac34\xp^2 -\frac12 \frac{m^2}{H^2} \bigg)   - \frac{1}{12} \xp \big(1+ \xp^2 \big) \ln(2\Lambda)    \nonumber\\
&+& \frac{1}{12} \xp \big(1+ \xp^2 \big) \bigg[ \frac{i \pi}{2} -\psi(1) + \psi(\frac12+\mu-\kappa) + \psi(\frac12-\mu-\kappa) \bigg] +\frac{1}{24} w(-1-\kappa) \nonumber\\
&-& \frac{1}{144} \bigg[\xi(-31+36\mu^2-13\xp^2) + 3i\lambda(-7+12\mu^2-4\xp^2)\bigg]\bigg\}\,.
\ee

Finally, summing up all the $\mathcal{G}_{i}$s and using $\kappa_{\lambda} = -\frac{ i \lambda}{2}\xp$ and $\mathcal{A}= -2\xp(1+\xp^2)$, we find the desired $\mathcal{G}_{\lambda}$ as 
\be\label{G-tot-lambda}
\mathcal{G}_{\lambda}(\Lambda) &=& - \lim_{\Lambda\rightarrow \infty}
 \bigg\{ \frac{\lambda}{9} \Lambda^{3} - \frac16 \xp  \Lambda^{2}  +  \frac{\lambda}{3} \Lambda \bigg( 1 - \frac34\xp^2 -\frac12 \frac{m^2}{H^2} \bigg)   - \frac{1}{12} \xp \big(1+ \xp^2 \big) \ln(2\Lambda)    \nonumber\\
 &+&  \frac{i\pi}{24}  \xp \big(1+ \xp^2 \big)  - \frac{1}{12} \xp (-\frac{25}{12} +\mu^2 - \frac{13}{12} \xp^2) +   \frac{\lambda}{18} \mu (4 - 4\mu^2 + 3\xp^2) \bigg( \frac{ e^{2i\pi\kl}}{\sin(2\pi\mu)}  + \frac{ \cos(2\pi\mu)}{\sin(2\pi\mu)}  \bigg)   \nonumber\\
 &+&  \frac{i}{24} \xp(1+\xp^2)  \bigg[   \frac{(e^{2i\kl\pi}+e^{-2i\mu\pi})}{\sin(2\mu\pi)} \psi(\frac12-\kl-\mu) -   \frac{(e^{2i\kl\pi}+e^{2i\mu\pi})}{\sin(2\mu\pi)} \psi(\frac12-\kl+\mu) \bigg]
  \bigg\}. \nonumber\\
\ee
As we see, it has divergent terms of the order $3$, $2$, $1$ and log of $\Lambda$. Summing over the polarization states, we find 
\be
\mathcal{G}(\Lambda) &=& - \lim_{\Lambda\rightarrow \infty}
 \bigg\{  - \frac13 \xp  \Lambda^{2} - \frac{1}{6} \xp \big(1+ \xp^2 \big) \ln(2\Lambda)  +  \frac{i\pi}{12}  \xp \big(1+ \xp^2 \big)  +\frac{1}{6} \xp (\frac{25}{12} - \mu^2 + \frac{13}{12} \xp^2)  \nonumber\\
 &+&  \frac{1}{24} \xp(1+\xp^2)  \bigg[   \frac{(e^{\xp\pi}+e^{2\lvert\mu\rvert\pi})}{\sinh(2\lvert\mu\rvert\pi)} \psi(\frac12+\frac{i}{2}\xp-\mu) -   \frac{(e^{\xp\pi}+e^{-2\lvert\mu\rvert\pi})}{\sinh(2\lvert\mu\rvert\pi)} \psi(\frac12+\frac{i}{2}\xp+\mu) \bigg]\nonumber\\
  &+&  \frac{1}{24} \xp(1+\xp^2)  \bigg[   \frac{(e^{-\xp\pi}+e^{2\lvert\mu\rvert\pi})}{\sinh(2\lvert\mu\rvert\pi)} \psi(\frac12-\frac{i}{2}\xp-\mu) -   \frac{(e^{-\xp\pi}+e^{-2\lvert\mu\rvert\pi})}{\sinh(2\lvert\mu\rvert\pi)} \psi(\frac12-\frac{i}{2}\xp+\mu) \bigg]
 \nonumber\\
 &+& \frac{\mu}{9}(4-4\mu^2 +3\xp^2)\frac{\sinh\xp \pi}{\sin2\mu\pi} \bigg\}.
\ee
Using the following relation between the polygamma functions (for $\mu=i\lvert\mu\rvert$)
\be
  && {\rm{Im}}\bigg[\frac{(e^{\xp\pi}+e^{2\lvert\mu\rvert\pi})}{\sinh(2\lvert\mu\rvert\pi)} \psi(\frac12+\frac{i}{2}\xp-\mu) -   \frac{(e^{\xp\pi}+e^{-2\lvert\mu\rvert\pi})}{\sinh(2\lvert\mu\rvert\pi)} \psi(\frac12+\frac{i}{2}\xp+\mu)  \nonumber\\  &&
   +  \frac{(e^{-\xp\pi}+e^{2\lvert\mu\rvert\pi})}{\sinh(2\lvert\mu\rvert\pi)} \psi(\frac12-\frac{i}{2}\xp-\mu) -  \frac{(e^{-\xp\pi}+e^{-2\lvert\mu\rvert\pi})}{\sinh(2\lvert\mu\rvert\pi)} \psi(\frac12-\frac{i}{2}\xp+\mu) \bigg] = -2\pi\,, \nonumber 
  \ee
we can further simplify $\mathcal{G}(\Lambda)$ to 
\be\label{G-tot}
\mathcal{G}(\Lambda) &=&  \lim_{\Lambda\rightarrow \infty}
 \bigg\{   \frac13 \xp  \Lambda^{2} + \frac{1}{6} \xp \big(1+ \xp^2 \big) \ln(2\Lambda)    - \frac{1}{6} \xp (\frac{25}{12} - \mu^2 + \frac{13}{12} \xp^2) - \frac{\mu}{9}(4-4\mu^2 +3\xp^2)\frac{\sinh\xp \pi}{\sin2\mu\pi} \nonumber\\
 &-&  \frac{1}{24} \xp(1+\xp^2) ~{\rm{Re}} \bigg[ \frac{(e^{\xp\pi}+e^{2\lvert\mu\rvert\pi})}{\sinh(2\lvert\mu\rvert\pi)} \psi(\frac12+\frac{i}{2}\xp-\mu) -   \frac{(e^{\xp\pi}+e^{-2\lvert\mu\rvert\pi})}{\sinh(2\lvert\mu\rvert\pi)} \psi(\frac12+\frac{i}{2}\xp+\mu) \nonumber\\
  &+&   \frac{(e^{-\xp\pi}+e^{2\lvert\mu\rvert\pi})}{\sinh(2\lvert\mu\rvert\pi)} \psi(\frac12-\frac{i}{2}\xp-\mu) -   \frac{(e^{-\xp\pi}+e^{-2\lvert\mu\rvert\pi})}{\sinh(2\lvert\mu\rvert\pi)} \psi(\frac12-\frac{i}{2}\xp+\mu) \bigg]
  \bigg\}.
\ee
The final relation is real and has divergent terms of the order $2$ and log of $\Lambda$.

\section{Adiabatic Subtraction and the regularized current}\label{Adiabatic Subtraction}

We use the adiabatic subtraction technique in curved QFT to remove the divergent terms in the current. 
Consider the WKB approximation form of the mode function
\be
\bar{u}_{\lambda,\bk}(\tau)=\frac{1}{(2\pi)^{\frac32}\sqrt{2W_{\lambda,\bk}}(\tau)} e^{- i \int^{\tau}_{-\infty} d\tilde{\tau} W_{\lambda,\bk}(\tilde\tau)}\,,
\ee
which is equal to the exact solution of the field equation when
\be\label{W}
W_{\lambda,\bk}^2 = \Omega^2_{\lambda,\bk} -\frac{a''}{a} +\frac34 \bigg(\frac{W'_{\lambda,\bk}}{W_{\lambda,\bk}}\bigg)^2 -\frac12 \frac{W''_{\lambda,\bk}}{W_{\lambda,\bk}}, 
\ee
in which
\be
 \Omega^2_{\lambda,\bk} = k^2-\lambda \xp k \mH +(\frac34 \xp^2 + \frac{m^2}{H^2}) \mH^2 . 
\ee
Notice that for all values of $k$, $\Omega^2_{\lambda,\bk}$ is positive definite, provided $m^2>0$. 

When $W_{\lambda,\bk}$ is real and positive, $\bar{u}_{\lambda,\bk}(\tau)$ corresponds to the (canonically normalized) positive frequency modes in the asymptotic past, i.e., the Wronskian is equal to $i(2\pi)^{-3}$. 

To use the adiabatic subtraction technique, here we define an adiabatic parameter $T^{-1}$ which in the limit $T \rightarrow0$ parametrizes an infinitely slow varying background geometry \cite{parker2009quantum}. We, then, assign a power of $T^{-1}$ to each time derivative in the last three terms in \eqref{W}. For the adiabatic subtraction approach, one needs to only expand $W_{\lambda,\bk}$ up to the $T^{-2}$ order 
\be\label{W-2}
W^2_{\lambda,\bk}  = \Omega_{\bk}^2 -\frac{a''}{a} +\frac34  \bigg(\frac{\Omega'_{\lambda,\bk}}{\Omega_{\lambda,\bk}}\bigg)^2 -\frac12 \frac{\Omega''_{\lambda,\bk}}{\Omega_{\lambda,\bk}} + \mathcal{O}(T^{-4}).
\ee
In computation of the current integral, we must keep terms up to $T^{-2}$ order and not any further. For more details on the adiabatic subtraction, see Ref. \cite{parker2009quantum}.

Using \eqref{W-2} in \eqref{curly-J} and expanding the integrand up to order $T^{-2}$, we have
\be
\mJ^{(2)}_{\rm{WKB}}(\tau,\vx) = \frac{ \ga H^3}{(2\pi)^2} \sum_{\lambda=\pm1} \mathcal{G}^{(2)}_{\lambda,\rm{WKB}},
\ee
where
\be 
\mathcal{G}^{(2)}_{\lambda,\rm{WKB}}= \lim_{\Lambda \rightarrow\infty} \int^{\Lambda}_{0} \x' d\x'\frac{k}{\Omega_{\lambda,\bk}} \big( -\frac13\lambda \x' + \frac12 \xp \big) \bigg[ 1 + \frac{a''}{2a \Omega^2_{\lambda,\bk}} -\frac38 \big(\frac{\Omega'_{\lambda,\bk}}{\Omega^2_{\lambda,\bk}}\big)^2 +\frac14 \frac{\Omega''_{\lambda,\bk}}{\Omega^3_{\lambda,\bk}}\bigg].\,\,\,
\ee
Dropping the slow-roll suppressed contributions, the above can be written as
\be 
\mathcal{G}^{(2)}_{\lambda,\rm{WKB}}&=& \lim_{\Lambda \rightarrow\infty} \int^{\Lambda}_{0} \x'^2 d\x'\frac{\mH}{\Omega_{\lambda,\bk}} \big( -\frac13\lambda \x' + \frac12 \xp \big) \bigg\{ 1 + \frac{\mH^2}{ \Omega^2_{\lambda,\bk}}  \nonumber\\
&+& \frac{\mH^4}{4\Omega^4_{\lambda,\bk}}\left[-\lambda \xp \x' +(\frac{9}{4}\xp^2+\frac{3m^2}{H^2})\right] -\frac58 \frac{\mH^6}{\Omega^6_{\lambda,\bk}}\left[-\frac{\lambda}{2}\xp \x' +(\frac34 \xp^2 +\frac{m^2}{H^2})\right]^2 \bigg\}\nonumber
\ee
and we have 
\be
&& \mathcal{G}^{(2)}_{\lambda,\rm{WKB}} = - \lim_{\Lambda\rightarrow \infty}  
 \bigg[ \frac{\lambda}{9} \Lambda^{3} - \frac16 \xp  \Lambda^{2}  +  \frac{\lambda}{3} \Lambda \big( 1 - \frac34\xp^2 -\frac12 \frac{m^2}{H^2} \big)   - \frac{1}{12} \xp \big(1+ \xp^2 \big) \ln(2\Lambda)    \nonumber\\
&+&\frac{\xp}{12}(1+\xp^2) \ln\big(\frac{\sqrt{9-4\mu^2}-\lambda \xp}{2}\big) + \frac{\xp}{12}\big(\frac{25}{12} - \mu^2 + \frac{13}{12}\xp^2 \big)  - \frac{\lambda}{12} \sqrt{1 -\frac49 \mu^2} \big(4\mu^2-3\xp^2+\frac12)   \nonumber\\
&-& \frac{\lambda}{8}\sqrt{9-4\mu^2} ~\xp^2  \frac{(27-12\mu^2-\frac{25}{9}\xp^2)}{(-9+4\mu^2+\xp^2)^2}  + \frac{\xp^3}{9} \frac{(\frac{13}{4}\xp^2 + 14\mu^2 -\frac{63}{2})}{(-9+4\mu^2+\xp^2)^2} \bigg].
\ee
The adiabatic subtraction quantity of each polarization state, $\mathcal{G}^{(2)}_{\lambda,\rm{WKB}}$, which is similar to $\mathcal{G}_{\lambda}$, contains divergences of the order $3$, $2$, $1$ and log of $\Lambda$. Summing over the polarization states, we obtain
\be\label{cG-WKB}
&& \mathcal{G}^{(2)}_{\rm{WKB}} = - \lim_{\Lambda\rightarrow \infty}  
 \bigg[ - \frac13 \xp  \Lambda^{2}   - \frac{1}{6} \xp \big(1+ \xp^2 \big) \ln(2\Lambda)  +\frac{\xp}{12}(1+\xp^2) \ln\big(\frac{9-4\mu^2 -\xp^2}{4}\big) \nonumber\\
&+& \frac{\xp}{6}\big(\frac{25}{12} - \mu^2 + \frac{13}{12}\xp^2 \big)   + \frac{\xp^3}{9} \frac{(\frac{13}{2}\xp^2 + 28 \mu^2 - 63)}{(-9+4\mu^2+\xp^2)^2} \bigg].
\ee
Note that $(9-4\mu^2 -\xp^2)$ which appears inside the log and also as a denominator, is positive definite and equal to $2(\xp^2+2\frac{m^2}{H^2})$.
As we see, the total subtraction quantity, $\mathcal{G}^{(2)}_{\rm{WKB}}$, has divergences of the order $2$ and log of $\Lambda$ which cancel the divergent terms in the total $\mathcal{G}$ given in \eqref{G-tot}.


\bibliographystyle{JHEP}
\bibliography{mybib}

\end{document}